\documentclass[amsmath,amssymb,11pt]{article}
\usepackage{jheppub2}
\pdfoutput=1
\usepackage{graphicx,epsfig}
\usepackage{float}
\usepackage{subfig}
\usepackage{subfloat}
\usepackage[utf8]{inputenc}
\usepackage{hyperref}
\usepackage{caption}
\usepackage{color}
\usepackage{overpic}
\usepackage[dvipsnames]{xcolor}
\usepackage{physics}
\usepackage[multiple]{footmisc}

\newcommand*{\affmark}[1][*]{\textsuperscript{#1}}

\newcommand{\beq}{\begin{equation}}

\newcommand{\eeq}{\end{equation}}

\usepackage{tikz}
\usetikzlibrary{positioning}
\usetikzlibrary{intersections}
\usetikzlibrary{fadings} 
\usetikzlibrary{arrows.meta} 
\usetikzlibrary{arrows}

\tikzfading[name=fade out,
inner color=transparent!0,
outer color=transparent!100]

\definecolor{cherryblossompink}{rgb}{1.0, 0.72, 0.77}
\definecolor{lightblue}{rgb}{0.68, 0.85, 0.9}

\usetikzlibrary{decorations.pathmorphing}
\usetikzlibrary{decorations.pathreplacing,decorations.markings}

\usetikzlibrary{backgrounds,automata}

\title{{\huge{Black holes in dS$_3$}}}
\subheader{\begin{flushright}
\texttt{IFT-UAM/CSIC-22-63\\
CPHT-RR045.062022}
\end{flushright}}

\author{Roberto Emparan,\affmark[1,2]}
\emailAdd{emparan@ub.edu}
\author{Juan F. Pedraza,\affmark[3]}
\emailAdd{j.pedraza@csic.es}
\author{Andrew Svesko,\affmark[4]}
\emailAdd{a.svesko@ucl.ac.uk}
\author{Marija Tomašević,\affmark[5]}
\emailAdd{marija.tomasevic@polytechnique.edu}
\author{and Manus R. Visser\affmark[6]}
\emailAdd{manus.visser@unige.ch}

\affiliation{\affmark[1]Institució Catalana de Recerca i Estudis Avançats (ICREA)\\
 Passeig Lluis Companys, 23, 08010 Barcelona, Spain\\
\affmark[2]Departament de Física Quàntica i Astrofísica and
  Institut de Ciències del Cosmos,\\
 Universitat de Barcelona, 08028 Barcelona, Spain\\
\affmark[3]Instituto de F\'isica Te\'orica UAM/CSIC\\
Calle Nicol\'as Cabrera 13-15, Cantoblanco, 28049 Madrid, Spain\\
\affmark[4]Department of Physics and Astronomy, University College London,\\
Gower Street, London, WC1E 6BT, United Kingdom\\
\affmark[5]Centre de Physique Théorique (CPHT), Ecole Polytechnique, \\
Bâtiment 6, Route de Saclay, 91128 Palaiseau, Cedex, France\\
\affmark[6]University of Geneva, Department of Theoretical Physics\\
24 quai Ernest-Ansermet, 1211 Gen\`{e}ve 4, Switzerland}

\abstract{In three-dimensional de Sitter space classical black holes do not exist, and the Schwarzschild-de Sitter solution  instead describes a conical defect with a single cosmological horizon. We argue that the quantum backreaction of conformal fields can generate a black hole horizon, leading to a three-dimensional quantum de Sitter black hole. Its size   can be as large as the cosmological horizon in a Nariai-type limit. We show explicitly how these solutions arise using braneworld holography, but also compare to a non-holographic, perturbative analysis of backreaction due to conformally coupled scalar fields in conical de Sitter space. We analyze the thermodynamics of this quantum black hole, revealing it behaves similarly to its classical four-dimensional counterpart, where the generalized entropy replaces the classical Bekenstein-Hawking entropy. We compute entropy deficits due to nucleating the three-dimensional black hole and revisit arguments for a possible matrix model description of dS spacetimes. Finally, we comment on the holographic dual description for dS spacetimes as seen from the braneworld perspective.
}

\begin{document}

\maketitle

\section{Introduction} \label{sec:intro}

\paragraph{Three-dimensional black holes, lost and found.} Lowering the dimensionality of spacetime simplifies the study of gravity, but often at a hefty price: black holes are wont to depart the scene. This follows from simple dimensional arguments. In three spacetime dimensions, which will be the focus of this article,  if we can only use Newton's constant $G_3$, then the presence of a massive object does not by itself introduce any length scale, since $G_3 M$ is a dimensionless quantity.\footnote{We always set the speed of light equal to one.} Therefore, there cannot be any black hole horizon solely determined by the mass of an object. Instead, the gravitational effect of a particle coupled to 
gravity becomes manifest only as a scale-free conical deficit \cite{Deser:1983tn}.

A cosmological constant can remedy this and allow black holes with a size proportional to the radius of three-dimensional Anti-de Sitter (AdS) space \cite{Banados:1992wn,Banados:1992gq}. However, although a length scale is necessary to have a horizon, it is not sufficient: some form of gravitational attraction is also needed. The tendency to collapse in AdS does it, but in de Sitter (dS) space the effect goes the other way around, and only a cosmological horizon, not a black hole, results from the cosmological length scale. Explicitly, in dS$_3$ with cosmological constant $\Lambda_3=1/R_3^2$, the geometry for a particle of mass $M$ at the pole is, in static coordinates \cite{Deser:1983nh},
\beq
ds^{2}=-f(r)dt^{2}+f^{-1}(r)dr^{2}+r^{2}d\phi^{2}\;,\qquad f(r)= 1-8G_{3}M-\frac{r^{2}}{R_{3}^{2}} \;.\label{eq:dS3conical}
\eeq
This contains a conical singularity at $r=0$, with deficit angle 
\begin{equation}
    \delta=2\pi\left(1-\sqrt{1-8G_3 M}\right)\,,
\label{eq:deficitangleintro}\end{equation}
and no black hole. In fact, since  constant time slices in dS$_3$ are two-spheres, globally there are two conical singularities in the maximal extension of \eqref{eq:dS3conical}, one at every pole of the two-sphere. 

A more subtle way of introducing a length scale is via quantum effects: with $\hbar\neq 0$, the Planck length \begin{equation}
    L_P=\hbar G_3
\end{equation}
makes an appearance.\footnote{In contrast, there is no notion of a three-dimensional Planck mass.} But even before knowing how this scale may enter to yield black holes, one may fear that it will fail to do so in a sensible way. If the black hole size must be proportional to the Planck length, then quantum gravitational effects might render the semi-classical description unreliable. To understand that this need not be so, note, first, that the presence of $\hbar G_3$ in this discussion does not immediately imply that quantum gravity must be important, but only that both gravitational and quantum effects are at play, \emph{e.g.}, with quantum fields coupled to classical gravity. When a large number $c\gg 1$ of these fields are introduced, the energy of their combined quantum effects, $\propto c\hbar$, may gravitate to give rise to a large semi-classical black hole horizon of radius $\sim G_3 c\hbar=c L_{P}\gg L_{P}$, near which quantum gravity effects would be relatively small. In the limit where $c\to\infty$ and $L_P\to 0$ with $c L_{P}$ fixed, such effects are absent, while the classical gravitational backreaction of the quantum fields remains finite. It is then conceivable that this backreaction results in a black hole of radius $\sim c L_{P}$.\footnote{In all dimensions, classical Einstein gravity
coupled to scale-invariant matter (quantum or classical) is itself a scale-invariant theory, such that in the setup described one can always choose units where $c L_{P}=1$.}\footnote{The large number of fields also lowers the cutoff energy scale of the quantum theory down to $1/(cL_P)$. The consequences in this context were discussed in \cite{Emparan:2002px} and will be reviewed later below.} Interestingly, its Bekenstein-Hawking entropy will then be $\sim c$, with no factors of $\hbar$, which indicates that it originates from microscopic one-loop effects in quantum field theory.

\paragraph{Quantum backreaction.} This mechanism was realized in \cite{Emparan:2002px} to obtain black holes in three-dimensional asymptotically locally flat space, as well as AdS$_3$ black holes with masses lower than the BTZ black holes (see also \cite{Emparan:2020znc}). 
Here we will employ it in three-dimensional de Sitter space. It can be convenient to envisage it as a two-step process. First, one solves for a quantum field in the spacetime \eqref{eq:dS3conical}.
The conical periodicity conditions give rise to a Casimir effect. For a free conformal scalar, we find that this results in a renormalized stress tensor of the form
\begin{equation}\label{casstress}
    \langle T^\mu{}_\nu\rangle =\frac{ F(M)}{8\pi r^3}\textrm{diag}(1,1,-2)\,,
\end{equation}
with $F(M)>0$. Therefore, the Casimir energy density in \eqref{casstress} is negative, but when, in the next step, we  compute its backreaction on the geometry, we find
\begin{equation}
    \delta g_{tt}=\frac{2 L_P F(M)}{r}>0\,,
\end{equation}
which means that the gravitational effect is attractive \cite{Souradeep:1992ia,Soleng:1993yh}.\footnote{Briefly, the reason is the following. A region of localized negative energy has a repulsive effect in its exterior, but the more one enters the region, the less repulsion is felt. As a result, at finite $r$ there is an effective attraction from the energy in \eqref{casstress}. We elaborate on this explanation in Appendix~\ref{app:negenergy}.
} Then, if a large number of fields are present, a semi-classical black hole horizon may appear. To prove this, the backreaction of the large number of fields must be non-linearly accounted for, that is, one must simultaneously solve the quantum field and the gravitational equations. The only framework that we know of where this can be consistently done in three or more dimensions is braneworld holography.

In this setup, classical dynamics in an $\text{AdS}_{d+1}$ bulk with a $d$-dimensional brane holographically encodes the quantum dynamics of the dual $d$-dimensional conformal field theory
coupled to a $d$-dimensional gravitational theory on the brane. In our context, the semi-classical Einstein equations in a four-dimensional AdS bulk are recast in the three-dimensional form
\beq
G_{\mu\nu}+\frac{1}{R_{3}^{2}}g_{\mu\nu}+\dots=8\pi G_{3}\langle T_{\mu\nu}\rangle\;,\label{eq:semiclassEin}
\eeq
where $g_{\mu\nu}$ is the metric induced on the brane, with curvature radius $R_{3}$ and Einstein tensor $G_{\mu\nu}$, and the dots denote higher curvature terms which can be systematically computed order by order \cite{Emparan:2020znc,Bueno:2022log}. These can be regarded, in dual terms, as induced by integrating out the holographic CFT degrees of freedom above the ultraviolet cutoff that the brane represents. The CFT below this cutoff gives rise to a renormalized $\langle T_{\mu\nu}\rangle$, with a large central charge $c\gg 1$ given by the AdS$_4$ radius in four-dimensional Planck units. Crucially, the bulk solution exactly encodes the quantum backreaction of the CFT on the three-dimensional geometry.

\paragraph{Quantum black hole in dS$_3$.} In this article, we apply this holographic approach to obtain black holes from quantum backreaction in three-dimensional de Sitter space. 
As in \cite{Emparan:2002px,Emparan:2020znc}, we use an exact solution of a black hole in an $\text{AdS}_{4}$ braneworld, 
but now with a brane with large enough tension 
such that
the effective cosmological constant on the brane is positive.\footnote{A black hole solution was found in \cite{deBuyl:2013ega} for a massive gravity theory in dS$_3$. Although the ``new massive gravity'' action of that article contains a term of the same form as the quadratic curvature term in our brane effective action, the two theories differ, and in particular our theory of gravity in dS$_3$ is not massive.} 

Using this solution, we find that   the holographic non-linear backreaction changes the conical geometry \eqref{eq:dS3conical} to have
\beq
f(r)= 1-8G_{3}M-\frac{r^{2}}{R_{3}^{2}}-\frac{2c L_{P} F(M)}{r} \;.\label{eq:qbhds3}
\eeq
It is easy to verify that the quantum $1/r$ term gives rise to a black hole horizon. This ``quantum Schwarzschild-dS$_{3}$'' (qSdS) solution is exact to all orders in the backreaction, in the planar limit of the CFT. The corrections are proportional to the central charge $c$ of the CFT, and the function $F(M)$ is now obtained by demanding regularity of the four-dimensional bulk. It differs from that of free scalars in \eqref{casstress}, where $c=1$, but the radial $1/r$ dependence (which is natural from the four-dimensional holographic perspective) and the tensorial structure of the corrections are the same in both cases.

Observe also that 
the physical range of masses in the conical spacetime \eqref{eq:dS3conical} is bounded above, 
\begin{equation}
    0< M < \frac1{8 G_3}\,,\label{eq:massrangev1}
\end{equation}
with the maximum mass reached when the conical deficit eats up all the space. The mass of the black holes in \eqref{eq:qbhds3} is also bounded above, but now the effect is due to the black hole horizon becoming as large as the cosmological horizon. This is a Nariai limit \cite{Nariai99,Ginsparg:1982rs} analogous to the one in Schwarzschild-de Sitter solutions in $d\geq 4$,  and it gives an upper mass bound that is lower (or equal) compared to \eqref{eq:dS3conical}. Since this limit is due to the appearance of the $1/r$ term in the metric, we expect that it is not exclusive of the holographic construction, but instead a generic property of quantum black holes in dS$_3$.

\paragraph{Thermodynamics.} A central focus of this article is  the   thermodynamics of the quantum black hole. This depends on the specific form of $F(M)$, so our results for the horizon entropies  are dependent on the holographic realization of the solution, but the form of the first laws below is expected to hold generically for semi-classical de Sitter gravity.

From the bulk perspective, the classical four-dimensional Bekenstein-Hawking entropy of each horizon is holographically understood to be the generalized entropy $S_{\text{gen}}^{(3)}$ in three dimensions \cite{Emparan:2006ni,Emparan:2020znc}.
The separation of the high- and low-energy CFT degrees of freedom also delimit the generalized entropy into the Wald entropy $S_{\text{Wald}}$ (accounting for higher curvature corrections), and entanglement entropy $S_{\text{out}}$ generated by the CFT living outside of each horizon,
\beq S_{\text{gen}}^{(3)}=S_{\text{Wald}}+S_{\text{out}}\;.\eeq
Further, each horizon of the quantum Schwarzschild-de Sitter black hole obeys a first law of thermodynamics, 
\beq dM=T_{h}dS^{(3)}_{\text{gen},h}\;,\quad dM=-T_{c}dS^{(3)}_{\text{gen},c}\;,\label{eq:firstlawsintro}\eeq
where $T_{h,c}=\frac{\kappa_{h,c}}{2\pi}$ refers to the  Gibbons-Hawking temperature of the two horizons, with surface gravities $\kappa_{h,c}$ defined with respect to the time translation Killing vector $\partial_t$, and $M$ is the mass. Importantly, the generalized entropy is an exact classical four-dimensional quantity, while the mass $M$, and temperatures $T_{c,h}$ are all three-dimensional quantities measured on the brane. Thence, as with the first law of the quantum BTZ black hole \cite{Emparan:2020znc}, the first laws (\ref{eq:firstlawsintro}) represent a non-trivial test of braneworld holography. Adding the two first laws yields
\beq 0=T_{h}dS_{\text{gen},h}^{(3)}+T_{c}dS_{\text{gen},c}^{(3)}\;,\label{eq:firstlawsumintro}\eeq
a three-dimensional analog of the semi-classical  first law  in $\text{dS}_{2}$ explored in \cite{Svesko:2022txo},\footnote{In the two-dimensional context the generalized entropy arises from including the 1-loop Polyakov action to describe the backreaction effects of a two-dimensional (not necessarily holographic) CFT.} and a semi-classical generalization of the usual first law of higher dimensional SdS black holes \cite{Gibbons:1977mu}. Notably, the first laws (\ref{eq:firstlawsintro}) and (\ref{eq:firstlawsumintro}) hold to all orders in backreaction and higher curvature corrections. \\

\noindent \textbf{Outline.} This article is structured as follows. In Sec.~\ref{sec:Qstresstensec} we consider a massless scalar field conformally coupled in a conical $\text{dS}_{3}$ background and derive the quantum corrected geometry perturbatively. In Sec.~\ref{sec:Cmetbranegrav} we describe the bulk $\text{AdS}_{4}$ geometry including a brane with a $\text{dS}_{3}$ slicing. We briefly review the gravitational theory induced on the brane, and uncover the semi-classical gravitational equations of motion, at least to second order in the strength of backreaction. Sec.~\ref{sec:qSdSBH} is devoted to finding the black hole solution localized on the brane, which is interpreted as the quantum three-dimensional  Schwarzschild-de Sitter black hole. We also detail the Nariai limit of the qSdS solution, which is nearly identical to the Nariai limit of the classical four-dimensional SdS black hole. In Sec.~\ref{sec:qSdSthermo} we analyze the thermodynamics of the quantum black hole. We find the three-dimensional thermodynamic quantities $S_{\text{gen}}^{(3)}$, $T$, and $M$ behave  similar to their classical four-dimensional counterparts. 
In Sec.~\ref{sec:entdeficit}, we compute  the entropy deficit between the \emph{generalized} entropies of the $\text{dS}_{3}$  and the qSdS$_3$ horizons surrounded by a CFT, which extend arguments hinting at a matrix model description of dS spacetimes and are also used to calculate the nucleation rate of a (quantum) black hole appearing in $\text{dS}_{3}$. In Sec.~\ref{sec:commentsdSholo} we comment on a possible realization of dS/CFT which naturally arises from holographic braneworlds. We conclude in Sec.~\ref{sec:disc}, where we outline multiple future research avenues. 

To keep this article self-contained we include multiple appendices. In App.~\ref{app:negenergy} we explain how the negative Casimir energy generated by a conical defect leads to an attractive gravitational potential. In App.~\ref{app:Qstressten} we provide the computational details of the quantum backreaction due to a massless conformally coupled scalar field in a $\text{dS}_{3}$ conical defect background. App.~\ref{app:bulkdualCFT} shows that in the limit of zero backreaction the $\text{AdS}_{4}$ bulk geometry is equal to the hyperbolic $\text{AdS}_{4}$ black hole upon a double Wick rotation. App.~\ref{app:susskind} provides computational details of the entropy deficit of quantum de Sitter black holes.
\section{Perturbative backreaction of quantum fields in conical dS$_3$}
\label{sec:Qstresstensec}

Consider a massless scalar field $\Phi$ conformally coupled to Einstein gravity in three dimensions,
\beq 
\begin{split}
I&= \frac{1}{16\pi G_{3}}\int d^{3}x\sqrt{-g}[R-2\Lambda]-\frac{1}{2}\int d^{3}x\sqrt{-g}\left[(\nabla\Phi)^{2}+\frac18 R\Phi^{2}\right]\;.
\end{split}
\label{eq:confmassfieldEHact}\eeq
We are interested in the renormalized stress-energy tensor of the field in the geometry \eqref{eq:dS3conical} of a conical defect in $\text{dS}_{3}$. It can be computed using a method analogous to that of the $\text{AdS}_{3}$ case in \cite{Steif:1993zv}, and here we only sketch it, deferring the details to App.~\ref{app:Qstressten}.

Generically, the Green's function for the scalar field equation in dS$_3$ is
\beq G(x,x')=\frac{1}{4\pi}\frac{1}{|x-x'|}+\frac{\lambda}{4\pi}\frac{1}{|x+x'|}\;,
\label{greens}\eeq
where $|x-x'|\equiv\sqrt{(x-x')^{a}(x-x')_{a}}$ is the chordal or geodesic distance between $x$ and $x'$ in the four-dimensional embedding space $\mathbb{R}^{2,2}$. 
The parameter $\lambda$ corresponds to different boundary conditions imposed on $G(x,x')$, namely `transparent' ($\lambda=0$), Neumann ($\lambda=1$), or Dirichlet ($\lambda=-1$) boundary conditions \cite{Avis:1977yn,Lifschytz:1993eb}. We will focus on the case of transparent boundary conditions. These are analogous to the definition in $\text{AdS}_{3}$, where they correspond to the case where the scalar field modes defined with respect to the time translation Killing vector are smooth on the Einstein static universe, obtained from an appropriate conformal transformation of $\text{dS}_{3}$. The holographic approach that we will employ later naturally selects these conditions too. 

If we consider that the conical spacetime is a $\mathbb{Z}_N$ orbifold, then the Green's function in it can be computed by summing over the $N$ images under the discrete action of $\partial_\phi$. With that, the renormalized quantum stress tensor $\langle T_{\mu\nu}\rangle$ can be derived by appropriately taking point-split derivatives, and a finite result is obtained after subtracting the `vacuum' term. After all this is done, we find a stress tensor of the form \eqref{casstress}, with
\beq
F(M)=\hbar\frac{\gamma^{3}}{4\sqrt{2}}\sum_{n=1}^{N-1}\frac{3+\cos(2\pi n\gamma)}{[1-\cos(2\pi n\gamma)]^{3/2}}\;,\label{eq:Tmunuads3app}
\eeq
where the parameter
\begin{equation}
\gamma=\sqrt{1-8G_3 M}    
\end{equation}
is related to the deficit angle (\ref{eq:deficitangleintro}) as $\delta=2\pi(1-\gamma)$.

We can now compute the gravitational effect of this stress-energy by coupling it to the Einstein equations
\beq
G^{\mu}_{\;\nu}+\frac{1}{R_{3}^{2}}\delta^{\mu}_{\;\nu} =8\pi G_{3}\langle T^{\mu}_{\;\nu}\rangle\;,\label{eq:semiclassEin2}
\eeq
and solving these perturbatively in $G_3 \hbar =L_P$ around the conical defect metric \eqref{eq:dS3conical}.  Analogous to the computation of quantum backreaction in conical $\text{AdS}_{3}$ \cite{Martinez:1996gn,Martinez:1996uv}, we consider a static and circularly symmetric background with the metric ansatz in circle-radius gauge, such that  $g_{\phi\phi}=r^2$,
\beq ds^{2}=-A(r)dt^{2}+B(r)dr^{2}+r^{2}d\phi^{2}\;.\eeq
The $tt$, $rr$ and $\phi\phi$ components of the semi-classical Einstein equations (\ref{eq:semiclassEin2}) are, respectively, 
\beq
\begin{split}
-\frac{B'}{2rB^{2}}+\frac{1}{R_{3}^{2}}&=\frac{L_{P}F(M)}{r^{3}}\;,\\
\frac{A'}{2r AB}+\frac{1}{R_{3}^{2}}&=\frac{L_{P}F(M)}{r^{3}}\;,\\
 \frac{2ABA''-AA'B'-BA'^{2}}{4A^{2}B^{2}}&+\frac{1}{R_{3}^{2}}=-2\frac{L_{P}F(M)}{r^{3}}\;.
\end{split}
\eeq
The general solution for $A$ and $B$ will depend on two integration constants, one of which is set to unity upon reparameterizing the time coordinate. 
The quantum effects must be regarded as a perturbation around the conical spacetime, so we must solve the equations as
\begin{align}
    A&=1-8G_{3}M-\frac{r^{2}}{R_{3}^{2}}-\gamma_{tt}(r)\,,\\
    B&=\frac1{1-8G_{3}M-\frac{r^{2}}{R_{3}^{2}}-\gamma_{rr}(r)}=\frac1{1-8G_{3}M-\frac{r^2}{R_{3}^2}}\left(1+\frac{\gamma_{rr}(r)}{1-8G_{3}M-\frac{r^2}{R_{3}^2}}+\mathcal{O}(L_P)^2\right)\,,
\end{align}
with $\gamma_{tt}(r)$ and $\gamma_{rr}(r)$ quantities of first order in $L_P$.
Solving the equations to this order gives
 \begin{equation}
     \gamma_{tt}=\gamma_{rr}=\frac{2 L_P F(M)}{r}\,.
 \end{equation}
 The fact that $\gamma_{tt}>0$ indicates an attractive gravitational effect, which suggest that a black hole horizon might form where $\gamma_{tt}\approx 1-8G_{3}M-r^2/R_3^2$. Of course, the perturbative nature of this solution does not entitle us to definitively reach this conclusion, but it is worth noticing that if we write the backreacted metric in the form
\beq ds^{2}=-f(r)dt^{2}+f^{-1}(r)dr^{2}+r^{2}d\phi^{2}\;,\quad f(r)= 1-8G_{3}M-\frac{r^{2}}{R_{3}^{2}}-\frac{2L_{P}F(M)}{r} \;,\label{eq:backreactgeom}\eeq
then the redshift factor $f(r)$ has the same form as in the four-dimensional Schwarzschild-de Sitter solution, which does indeed have a black hole horizon. The interpretation of terms, however, is different: the $1/r$ term, which in four dimensions would be associated to the mass, here is due to quantum corrections, while the three-dimensional mass is given by the constant terms. This four-dimensional resemblance of the quantum-corrected geometry is obscure in this setup, but it will become natural within the holographic construction.


\section{de Sitter braneworld in AdS$_4$} \label{sec:Cmetbranegrav}

We turn now to the approach of braneworld holography. We shall start by reviewing the construction of a de Sitter brane in an AdS$_4$ bulk.

Its  main features can be conveniently understood starting from the metric of the  Rindler-AdS$_4$ spacetime, namely
\beq\label{eq:rindlerads4}
ds^{2}=-\left(\frac{\rho^{2}}{\ell_{4}^{2}}-1\right)dt_{R}^{2}+\frac{d\rho^{2}}{\frac{\rho^{2}}{\ell_{4}^{2}}-1}+\rho^{2}(d\vartheta^{2}+\sinh^{2}\vartheta\,d\phi^{2})\;.
\eeq
Orbits of $\partial_{t_R}$ are trajectories of uniform acceleration, and the surface $\rho=\ell_4$ is a non-compact, acceleration horizon, such that $\text{AdS}_{4}$ is not globally covered in these coordinates. With respect to the canonically normalized time $t_R$, the horizon has a temperature
\begin{equation}\label{TRAdS4}
    T_R=\frac1{2\pi\ell_4}\,.
\end{equation}
Let us now rewrite this spacetime after changing spatial coordinates as
\beq \frac{\hat{r}^{2}}{R_{3}^{2}}=\frac{\rho^{2}\sinh^{2}\vartheta}{\rho^{2}\cosh^{2}\vartheta-\ell_{4}^{2}}\;,\quad \cosh\sigma=\frac{\rho}{\ell_{4}}\cosh\vartheta\;,\eeq
and (as a matter of mere convenience, which does not change the patches that are covered)
\beq\label{resct}
\frac{t}{R_{3}}= \frac{t_{R}}{\ell_{4}}\,.
\eeq
The metric becomes
\beq ds^{2}=\ell_{4}^{2}d\sigma^{2}+\frac{\ell_{4}^{2}}{R_{3}^{2}}\sinh^{2}\sigma\left[-\left(1-\frac{\hat{r}^{2}}{R_{3}^{2}}\right)dt^{2}+\left(1-\frac{\hat{r}^{2}}{R_{3}^{2}}\right)^{-1}d\hat{r}^{2}+\hat{r}^{2}d\phi^{2}\right]\;.\label{eq:lineelementds3sec}\eeq
We see that sections of constant $\sigma$ yield $\text{dS}_{3}$ in static-patch coordinates with radius given by $\ell_{4}\sinh\sigma$. 
The acceleration horizon in \eqref{eq:rindlerads4} is now at $\hat{r}=R_{3}$, which, as $\sigma$ varies between $0$ and $\infty$, traces out the cosmological horizons of the $\text{dS}_{3}$ sections. That is, the bulk Rindler horizon induces a  cosmological horizon on the  $\text{dS}_{3}$ slices of constant $\sigma$. 

The bulk acceleration horizon is non-compact since it extends all the way to the asymptotic boundary at $\sigma\to\infty$. It will become a compact horizon if we make our universe compact by introducing a positive-tension brane
at
\begin{equation}\label{sigmab}
    \sinh\sigma_b=\frac{R_3}{\ell_4}\,,
\end{equation}
which excludes all of the region $\sigma>\sigma_b$. If the brane action is purely tensional,
\beq I_{\text{brane}}=-\tau\int d^{3}x\sqrt{-h}\;, \label{eq:braneact}\eeq
then the Israel junction conditions (i.e., the equations of motion of the brane) demand that the tension be
\begin{equation}
    \tau=\frac{\cosh\sigma_b}{2\pi G_4 R_3}=\frac1{2\pi G_4}\sqrt{\frac1{R_3^2}+\frac1{\ell_4^2}}\,.\label{tauR3ell4}
\end{equation}
For later convenience, instead of the tension $\tau$ we will use the associated length scale
\begin{equation}\label{elltau}
    \ell=\frac1{2\pi G_4 \tau}\,,
\end{equation}
such that $\cosh\sigma_b=R_3/\ell$. Eq.~\eqref{tauR3ell4} becomes
\begin{equation}\label{ellR3}
    \frac1{\ell^2}=\frac1{R_3^2}+\frac1{\ell_4^2}\,.
\end{equation}
Observe that $\ell<\ell_4$, i.e., for fixed $\ell_4$ the brane tension is bounded below in order for $R_3^2>0$. Branes with lower tension, $\ell=\ell_4$ and $\ell>\ell_4$, would have Minkowski$_3$ and AdS$_3$ worldvolumes, respectively.

\begin{figure}[t]
\begin{center}
\includegraphics[width=.235\textwidth]{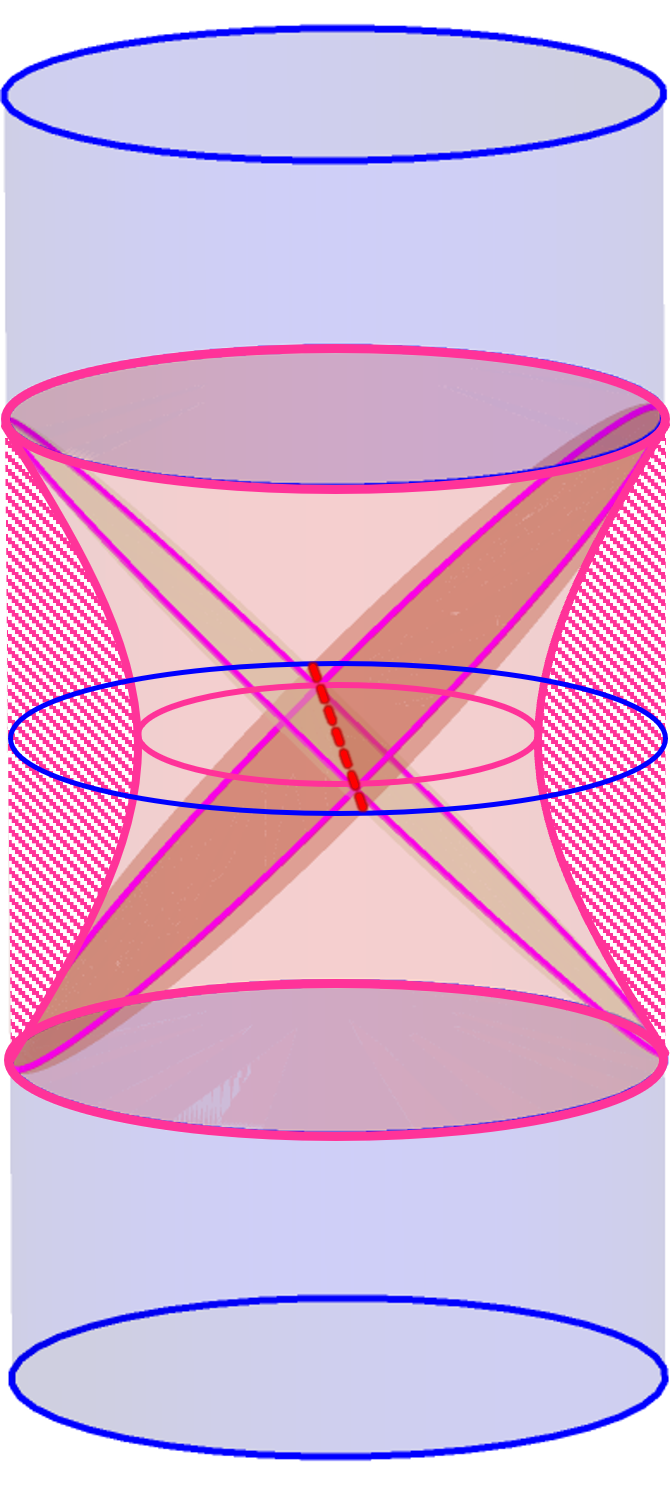}
\end{center}
\caption{\small  Bulk $\text{AdS}_{4}$ with a de Sitter$_3$ brane. The brane is represented as a (magenta) hyperboloid ($\sigma=\sigma_b$ in \eqref{eq:lineelementds3sec}), and gravity is induced on it from integrating out the UV degrees of freedom of the $\text{CFT}_{3}$ excluded by the brane (dashed magenta region). One performs surgery by gluing two copies of the region $\sigma<\sigma_b$ along the two-sided brane. The brane is following an accelerating trajectory, and the bulk acceleration horizons give rise to cosmological horizons on the induced dS geometry. The red dashed line is the bifurcation surface of the Rindler-AdS horizons. When the brane is introduced, it becomes a compact surface of finite area \eqref{compacthor}.}
\label{fig:AdSwithdSbrane} 
\end{figure}

The geometry induced on the brane is that of dS$_3$ with radius $R_3$, and the area of the horizon in the bulk is now finite, with a corresponding finite entropy
\begin{equation}
    S_{\text{BH}}^{(4)}=\frac{A_H}{4G_4}=\frac{\pi \ell_4^2}{2G_4}\left(\cosh\sigma_b-1\right)=\frac{\pi }{2G_4}\frac{R_3^2\ell}{R_3+\ell}\,.\label{compacthor}
\end{equation}
An illustration of the resulting spacetime is presented in Fig.~\ref{fig:AdSwithdSbrane}. 

\subsubsection*{Induced gravity theory}

So far we are considering the spacetime and the brane within it in a four-dimensional interpretation. However, in these braneworlds, a mode of the bulk graviton is localized near the brane \cite{Randall:1999ee,Randall:1999vf}, reproducing a three-dimensional theory of gravity there; in addition, using holography, the remaining bulk graviton modes can be described in terms of a dual three-dimensional CFT. As a result, the entire four-dimensional setup admits a purely three-dimensional description as a classical gravitational theory coupled to a quantum CFT.

The effective theory of gravity on the brane can be regarded as being induced by integrating out the  holographic UV degrees of freedom of the dual $\text{CFT}_{3}$ down to an energy cutoff of $1/\ell$ \cite{deHaro:2000wj}. 
The derivation given in \cite{Emparan:2020znc} for AdS$_3$ branes is also valid for dS$_3$ branes, so we skip directly to the result of the integration, namely (see also \cite{Chen:2020uac,Bueno:2022log})
\beq 
\begin{split}
I_{\text{ind}}&=\frac{\ell_{4}}{8\pi G_{4}}\int d^{3}x\sqrt{-h}\left[\frac4{\ell_4}\left(\frac1{\ell}-\frac1{\ell_{4}}\right)+\tilde{R}+\ell_{4}^{2}\left(\frac{3}{8}\tilde{R}^{2}-\tilde{R}_{\mu\nu}^{2}\right)+...\right],
\end{split}\label{indact}
\eeq
where $\tilde{g}_{\mu\nu}=h_{\mu\nu}$ and $\tilde{R}_{\mu\nu}$ are the metric induced on the brane and its Ricci curvature.
From here we identify the effective three-dimensional Newton's constant as
\beq G_{3}\equiv\frac{G_{4}}{2\ell_{4}}\;,\label{eq:G3}\eeq
and the effective three-dimensional cosmological constant,
\beq
\frac{2}{L_{3}^{2}}\equiv \frac4{\ell_4}\left(\frac1{\ell}-\frac1{\ell_{4}}\right)\;.\label{eq:branecc}\eeq
The higher curvature terms in the action are multiplied by higher powers of $\ell_4$, which plays the role of the cutoff length scale of the effective three-dimensional theory. More precisely, in order for the effective three-dimensional theory to be valid we must have
\begin{equation}
    \ell_4\ll L_3\,,
\end{equation}
which, using \eqref{ellR3}, means that
\begin{equation}
    \ell \sim \ell_4\ll R_3\,.
\end{equation}
From \eqref{sigmab} we see that this requires that the brane be close to the boundary, $\sigma_b\gg 1$.
We will often regard $\ell/R_3$ as the small expansion parameter of the effective theory.
Observe that $R_3$, which is defined as the physical curvature radius of the brane, is not exactly the same as $L_3$ due to the higher curvature terms. For small $\ell$ we find
\beq
\begin{split}
 \frac{1}{L_{3}^{2}}=\frac{2}{\ell_{4}^{2}}\left(\frac{\ell_{4}}{\ell}-1\right)\approx \frac{1}{R_{3}^{2}}-\frac{\ell^{2}}{4R_{3}^{4}}+\mathcal{O}(\ell^{4}/R_{3}^{6})\;.
\end{split}
\label{eq:approx}\eeq

The complete three-dimensional effective action $I$ on the brane is the sum of the induced gravity action $I_{\text{ind}}$ and the action $I_{\text{CFT}}$ of the boundary $\text{CFT}_{3}$, determined holographically by the bulk. Using (\ref{eq:approx}), to leading order we have
\beq I=\frac{1}{16\pi G_{3}}\int d^{3}x\sqrt{-h}\left[\tilde{R}-\frac{2}{R_{3}^{2}}+\ell^{2}\left(\frac{3}{8}\tilde{R}^{2}-\tilde{R}^{2}_{\mu\nu}\right)+...\right]+I_{\text{CFT}}\;,\label{eq:totalact}\eeq
where the ellipsis denotes terms of order $\mathcal{O}(\ell^{4})$ and higher. Further, following \cite{Emparan:2020znc}, we normalize the central charge $c$ of the boundary CFT as\footnote{Here we are working in units where $\hbar=1$. To recover factors of $\hbar$, one need only replace $c\to c\hbar$.}
\beq c=\frac{\ell_{4}^{2}}{G_{4}}\quad\Rightarrow\quad 2cG_{3}=\ell_{4}\approx \ell\left(1+\frac{\ell^{2}}{2R_{3}^{2}}+\frac{3}{8}\frac{\ell^{4}}{R_{3}^{4}}+...\right)\;.\label{eq:usefulapprox}\eeq
Therefore, for fixed $c$, as $\ell\to0$ gravity on the brane becomes weak ($G_{3}\to0$) such that there is no backreaction due to the CFT. This limit looks singular from the viewpoint of the bulk, since if we want to keep $R_3$ finite then we must take $\ell_4\to 0$. In the naive way of taking this limit we are not only eliminating the backreaction, but removing the CFT$_3$ altogether by removing the bulk.  If instead we take the limit $\ell_4\to 0$ while rescaling the bulk metric by a factor $\ell_4^2$,  then the brane is pushed to the boundary and gravitational dynamics on the brane is turned off, while still keeping a non-trivial state of the non-backreacting CFT$_3$. This limit is described in App.~\ref{app:bulkdualCFT}. Note also that, since the three-dimensional Planck length is $L_P=G_3$ (with $\hbar=1$), we can write \eqref{eq:usefulapprox}
as
\beq \ell= 2cL_P\left(1+\mathcal{O}\left(\frac{cL_P}{R_{3}}\right)^{2}\right)\;,
\eeq
which will often be useful.


According to the holographic dictionary, the induced metric on the brane solves the semi-classical gravitational equations \cite{Emparan:2020znc}
\begin{align} \label{eq:semiclasseom}
&8\pi G_{3}\langle T^{\text{CFT}}_{\alpha\beta}\rangle=\tilde{G}_{\alpha\beta}+\frac{h_{\alpha\beta}}{L_{3}^{2}}\\
&+\ell^{2}\biggr[4\tilde{R}_{\alpha}^{\;\gamma}\tilde{R}_{\beta\gamma}-\frac{9}{4}\tilde{R}\tilde{R}_{\alpha\beta}-\Box \tilde{R}_{\alpha\beta}+\frac{1}{4}\nabla_{\alpha}\nabla_{\beta}\tilde{R}+\frac{1}{2}h_{\alpha\beta}\left(\frac{13}{8}\tilde{R}^{2}-3\tilde{R}_{\gamma\delta}^{2}+\frac{1}{2}\Box\tilde{R}\right)\biggr]+ ...\;, \nonumber
\end{align}
where the CFT stress-energy tensor sources the effective three-dimensional gravity theory. With quantum backreaction accounted for by $\langle T_{\alpha\beta}^{\text{CFT}}\rangle$, black hole solutions to this model are quantum-corrected black holes \cite{Emparan:2002px}, such that the classical dynamics of the bulk four-dimensional Einstein theory encodes the quantum dynamics of the dual three-dimensional effective theory.

\section{Quantum Schwarzschild-de Sitter black hole} \label{sec:qSdSBH}


Now we introduce one of the main elements in this work: an exact solution to the bulk theory that describes a black hole localized on a dS$_3$ brane. According to our previous discussion, its dual interpretation is as a black hole solution to the semiclassical Einstein equations \eqref{eq:semiclassEin}, with the backreaction included exactly, to leading order in the large-$c$ expansion of the quantum CFT$_3$. Therefore, we can claim that the solution does describe a quantum black hole in dS$_3$. 

Our discussion follows a similar analysis presented in \cite{Emparan:2002px} for an AdS$_3$ brane, which led to the quantum BTZ (qBTZ) black hole, but there are some key differences worth highlighting.

\subsection{$\text{AdS}_{4}$ C-metric}

We begin with the following solution to Einstein's equation with negative cosmological constant, which is a particular case of the Plebanski-Demianski type-D solutions \cite{Plebanski:1976gy} 
\beq ds^{2}=\frac{\ell^{2}}{(\ell+xr)^{2}}\left[-H(r)dt^{2}+\frac{dr^{2}}{H(r)}+r^{2}\left(\frac{dx^{2}}{G(x)}+G(x)d\phi^{2}\right)\right]\;,\label{eq:AdS4Ccoord}\eeq
where the metric functions $H(r)$ and $G(x)$ are given by
\beq H(r)= 1-\frac{r^{2}}{R_{3}^{2}}-\frac{\mu\ell}{r} \;,\qquad G(x)=1-x^{2}-\mu x^{3}\;.\label{eq:Hrfunc}\eeq
Our conventions largely follow \cite{Emparan:2020znc}, however, we have selected $\kappa=+1$ and set $\ell_{3}^{2}=-R_{3}^{2}$ such that the brane we introduce is a $\text{dS}_{3}$ brane of radius $R_{3}$. 

Before introducing the brane, these solutions are well known to describe accelerating black holes in AdS$_4$ \cite{Emparan:1999wa,Emparan:1999fd}. The parameter $\mu>0$ would be interpreted as related to the mass of the four-dimensional black hole, but we will see that for us the interpretation is different. The parameter $\ell>0$ is the inverse of the acceleration of the black hole, $A=1/\ell$, but it is actually the same as the tension length we introduced before: the relation \eqref{ellR3} also holds when the bulk metric satisfies $R_{AB}=-(3/\ell_{4}^{2})g_{AB}$. The idea is that in our brane construction the tension of the brane provides the acceleration of the black hole that is attached to it, and we take large enough tension that there is an acceleration horizon in the bulk.\footnote{The regimes where the tension and the acceleration are small, without an acceleration horizon, are appropriate for AdS$_3$ branes, as in \cite{Emparan:1999fd,Emparan:2020znc}. Minkowski branes are obtained in the critical case \cite{Emparan:1999wa}.}

\begin{figure}[t]
\begin{center}
\includegraphics[width=.9\textwidth]{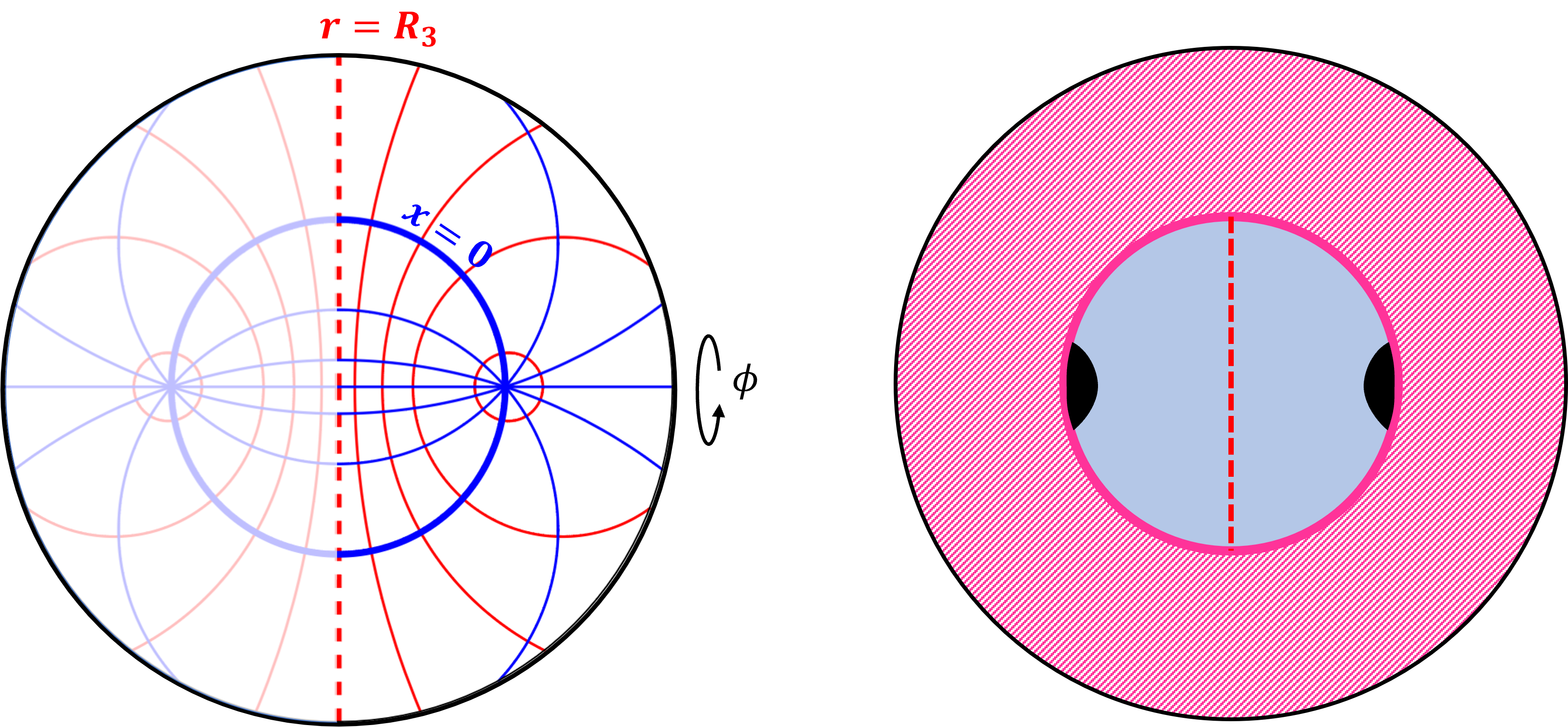}
\end{center}
\caption{\small \textbf{Left}: AdS$_4$ C-metric \eqref{eq:AdS4Ccoord} with $\mu=0$, in $(r,x)$ coordinates in a slice at $t=0$ and constant $\phi$.
Lines of constant $x$ are blue arcs; lines of
constant $r$ are red arcs (full circles for $0 < r <\ell$). The thick blue circle $x=0$ is where we place the dS$_3$ brane; its interior is $0<x\leq1$, with $x=1$ the $\phi$ axis of rotation. The exterior region $x<0$ is excluded in the braneworld construction. The vertical red dashed line is the horizon at $r=R_3$. Its intersection with the brane yields a dS$_3$ cosmological horizon.  The coordinates only cover half of the disk, with the other half being obtained through analytic continuation. \textbf{Right}: construction of black holes on a dS$_3$ braneworld when $\mu>0$. The dashed magenta region $x<0$ is excluded. The black hole horizon and the cosmological horizon are at constant $r$.}
\label{fig:ccoords} 
\end{figure}

The relation between our earlier brane construction and the metric \eqref{eq:AdS4Ccoord} becomes apparent if we consider the case when $\mu=0$ and perform the coordinate transformation
\beq \sinh \sigma =\frac{R_{3}}{\ell_{4}}\frac{\sqrt{1-x^{2}r^{2}/R_{3}^{2}}}{|1+rx/\ell|}\;,\qquad \hat{r}=r\sqrt{\frac{1-x^{2}}{1-x^{2}r^{2}/R_{3}^{2}}}\;.
\eeq
Then the metric (\ref{eq:AdS4Ccoord}) becomes the same as that of Rindler-AdS, eq.~\eqref{eq:lineelementds3sec} (see the left Fig.~\ref{fig:ccoords} for a representation of the $(r,x)$ coordinates).


Each zero of the function $G(x)$ corresponds to an axis for the rotation symmetry, with possible conical singularities lying there. Particularly, for a range of values of $\mu$, there will be three distinct zeros to $G(x)$, denoted by $\{x_{0},x_{1},x_{2}\}$, with each zero leading to a distinct conical singularity. One of the conical singularities can be removed via the identification, 
\beq \phi\sim \phi+\frac{4\pi}{|G'(x_{i})|}\;,\eeq
where $x_{i}$ is one of the select zeros. Once the period of $\phi$ has been fixed in this way, say at $x=x_{1}$, then $\phi$ cannot be readjusted to eliminate the remaining singularities at $x=x_{0},x_{2}$.

Not all of these zeroes of $G(x)$ will be relevant for our discussion, though. We are interested in introducing a brane, which in the case $\mu=0$ of empty AdS$_4$ is at \eqref{sigmab}, and in the coordinates of \eqref{eq:AdS4Ccoord} corresponds to $x=0$. 
A feature of the AdS$_4$ C-metric that makes it especially suitable for braneworld constructions is that,  when $\mu>0$, the surface $x=0$ also satisfies the Israel junction conditions for a brane with tension given by \eqref{elltau}. Thus, the braneworld we seek is obtained by placing such a brane at $x=0$, and keeping only the 
\begin{equation}
    x>0
\end{equation}
portion of the bulk geometry. The metric induced on the brane located at $x=0$ will be 
\beq\label{indmet0}
ds^{2}|_{x=0}=-\left(1-\frac{\mu\ell}{r}-\frac{r^{2}}{R_{3}^{2}}\right)dt^{2}+\left(1-\frac{\mu\ell}{r}-\frac{r^{2}}{R_{3}^{2}}\right)^{-1}dr^{2}+r^{2}d\phi^{2}\;,
\eeq
which is strongly reminiscent of the quantum-backreacted geometry \eqref{eq:backreactgeom} that we have encountered before. However, we note two differences. The first one is very significant: the solution \eqref{indmet0} is exact, so, unlike in \eqref{eq:backreactgeom}, the term $\propto 1/r$ in the metric coefficients does not need to be treated as a perturbative correction (even though it may be small). The second one is that it is unclear what are the physical parameters of the solution in \eqref{indmet0}: what is the three-dimensional mass $M$, and how is $\mu$ related to the quantum corrections $L_P F(M)$? For this purpose, we must analyze the global properties of the bulk solution.

\subsection*{Bulk regularity}

\begin{figure}[t]
\begin{center}
\includegraphics[width=.3\textwidth]{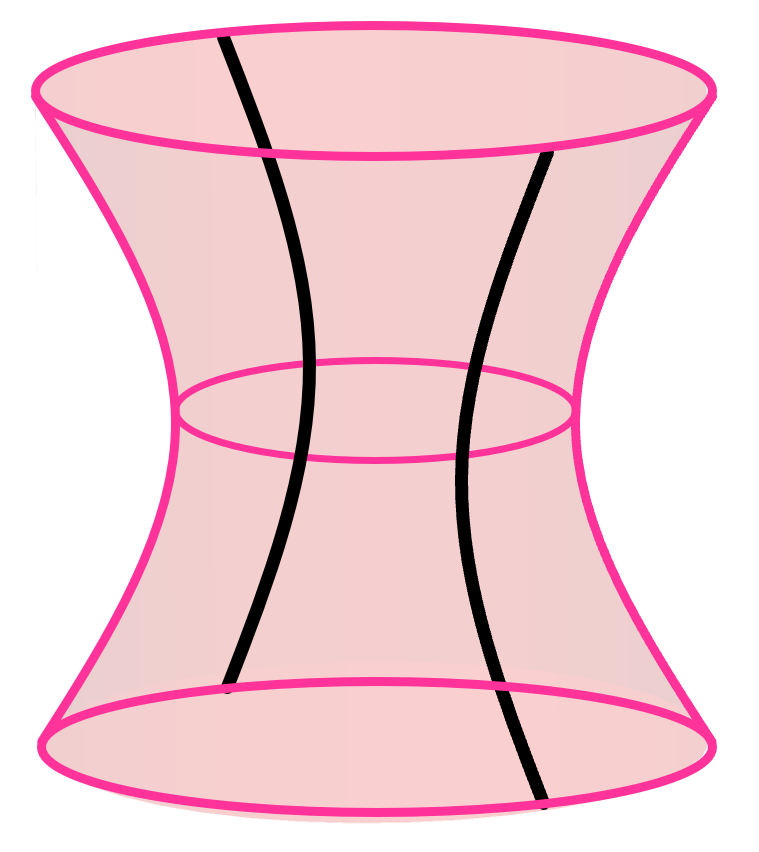}
\end{center}
\caption{\small Illustration of the bulk braneworld from the AdS$_4$ C-metric. The brane is placed at $x=0$ (magenta hyperboloid surface). The black lines denote $r=0$ and depict the worldlines of the accelerating black holes described by the C-metric. The AdS interior of the hyperboloid is kept, and the double-sided brane is glued to another copy of it.}
\label{fig:dsbranesid} 
\end{figure}

Black holes arise in the AdS$_4$ C-metric (\ref{eq:AdS4Ccoord}) when $\mu\neq0$ (see the right Fig.~\ref{fig:ccoords} and Fig.~\ref{fig:dsbranesid}).  Whether we find a black hole horizon depends on the nature of the roots of the functions $H(r)$ and $G(x)$, where the roots of $H(r)$ correspond to the Killing horizons generated by the time translation Killing vector $\partial_{t}$. We desire for positive roots of $H(r)$, if we are to describe physically acceptable horizons. Meanwhile, real roots of $G(x)$ correspond to symmetry axes of the Killing vector $\partial_{\phi}$, and they characterize the geometry of the horizon in the bulk. For instance, a surface of constant $r$ with $0\leq x\leq x_1$ is a (distorted) half-sphere with disk topology.

Our strategy is to first consider the roots of $G(x)$, where we look for at least one real root. This can be established by specifying a range for the parameter $\mu$ \cite{Emparan:1999wa,Emparan:1999fd}. When $\mu>0$ there will be one positive root, denoted by $x_{1}$, and we restrict the range of $x$ such that $0\leq x\leq x_{1}$. In this way, there is a single conical singularity at $x=x_{1}$, which may be removed by a proper identification of the period for $\phi$, while the remaining conical singularities simply will be absent via the spacetime surgery of introducing a brane at $x=0$. Following \cite{Emparan:2020znc}, it behooves us to consider the root $x_{1}$ as a parameter, while  $\mu$ is ``derived". That is,  from $G(x_{1})=0$ we have
\beq \mu=\frac{1-x_{1}^{2}}{x^{3}_{1}}\;,\label{eq:defmu}\eeq
with $x_{1}\in(0,1]$. We see $\mu$ will monotonically decrease from $+\infty$ to zero, where $\mu=0$ coincides with $x_{1}=1$. Later we will see that when $\ell\neq 0$ the allowed value of $\mu$ will be limited above if we want to have a regular black hole horizon.

The conical singularity at $x=x_{1}$ is removed via the identification
\beq \phi\sim\phi+\Delta\phi\;,\qquad \Delta\phi=\frac{4\pi}{|G'(x_{1})|}=\frac{4\pi x_{1}}{3-x_{1}^{2}}\;.\label{eq:conicaldef}\eeq
We see for the range of $x_{1}$, the function $G'(x_{1})=-\frac{3-x_{1}^{2}}{x_{1}}<0$, and that $\Delta\phi$ is independent of $\ell$ and $R_{3}$.  Moreover, $\Delta\phi$ grows monotonically from $0$ to $2\pi$.

\subsection{Black hole on the brane}

 
We now understand that, although the metric \eqref{indmet0} has the form of asymptotically $\text{dS}_{3}$ space, the $\phi$ coordinate is not identified with $2\pi$ but rather $\Delta\phi$ (\ref{eq:conicaldef}). Then there is a conical deficit angle which signals the effect of a mass. To make it manifest, we change to canonical coordinates $(t,r,\phi)\to(\bar{t},\bar{r},\bar{\phi})$ via
\beq t=\eta\bar{t}\;,\qquad r=\frac{\bar{r}}{\eta}\;,\qquad \phi=\eta\bar{\phi}\;,\eeq
where 
\begin{equation}
    \eta\equiv \frac{\Delta\phi}{2\pi}=\frac{2x_1}{3-x_1^2}
\end{equation}
and
\beq ds^{2}|_{x=0}=-\left(\eta^{2}-\frac{\mu\ell\eta^{3}}{\bar{r}}- \frac{\bar{r}^{2}}{R_{3}^{2}}\right)d\bar{t}^{2}+\left(\eta^{2}-\frac{\mu\ell\eta^{3}}{\bar{r}}- \frac{\bar{r}^{2}}{R_{3}^{2}}\right)^{-1}d\bar{r}^{2}+\bar{r}^{2}d\bar{\phi}^{2}\;.\label{eq:bhonbranev2}\eeq
Since now $\bar\phi\sim\bar\phi+2\pi$, we can identify $\eta^2$ with $1-8G_3 M$ and thus obtain the mass of the solution. Actually, it is convenient to perform the identification as 
\beq 8\mathcal{G}_{3}M=1-\eta^2=1-\frac{4x_{1}^{2}}{(3-x_{1}^{2})^{2}}\;,\qquad \mathcal{G}_{3}\equiv \frac{\ell_{4}}{\ell}G_{3}\;,\label{eq:Massx1}\eeq
where we have introduced the renormalized Newton's constant $\mathcal{G}_{3}$, following \cite{Emparan:2020znc}. This takes into account the modifications in the definition of the mass due to higher curvature corrections in the effective gravitational theory, which are then encoded in a `renormalized' Newton's constant $\mathcal{G}_{3}$ \cite{Cremonini:2009ih}.\footnote{We point out the relation $\mathcal{G}_{3}\equiv \frac{\ell_{4}}{\ell}G_{3}$ is assumed to hold for all orders in $\ell$ and may be interpreted as a resummation of higher curvature corrections to the mass at all orders in $\ell$.}

By virtue of (\ref{eq:defmu}) and (\ref{eq:conicaldef}), we can respectively replace $\mu$ and $\eta$ with  rational polynomials of $x_{1}$. Explicitly, the function $H(\bar{r})$ becomes
\beq H(\bar{r})=\frac{4x_{1}^{2}}{(3-x_{1}^{2})^{2}}-\frac{8\ell}{\bar{r}}\frac{1-x_{1}^{2}}{(3-x_{1}^{2})^{3}}-\frac{\bar{r}^{2}}{R_{3}^{2}}\;.\eeq
We have related the first term to a mass $M$ and a renormalized Newton's constant~$\mathcal{G}_{3}$; the additional $\ell/\bar{r}$ term characterizes quantum corrections to the black hole. It vanishes when $\ell\to 0$, which is the limit in which the gravitational effects of the CFT are suppressed, and indeed we recover a classical conical defect in dS$_3$. For finite $\ell$, however, the backreaction leads to a $1/\bar{r}$ (quantum) correction to $\text{SdS}_{3}$. 
Given the relation between $x_{1}$ and $M$ in (\ref{eq:Massx1}), we naturally interpret the second term in $H(\bar{r})$ as a function of the mass $F(M)$,
\beq F(M)\equiv \frac{8(1-x_{1}^{2})}{(3-x_{1}^{2})^{3}}\;.\label{eq:FMx1}\eeq
Altogether then, we recast the metric on the brane (\ref{eq:bhonbranev2}) as
\beq ds^{2}|_{x=0}=-H(\bar{r})d\bar{t}^{2}+H^{-1}(\bar{r})d\bar{r}^{2}+\bar{r}^{2}d\bar{\phi}^{2}\;, \quad H(\bar{r})=1-8\mathcal{G}_{3}M-\frac{\ell F(M)}{\bar{r}}-\frac{\bar{r}^{2}}{R_{3}^{2}}\;.\label{eq:qdS3}\eeq


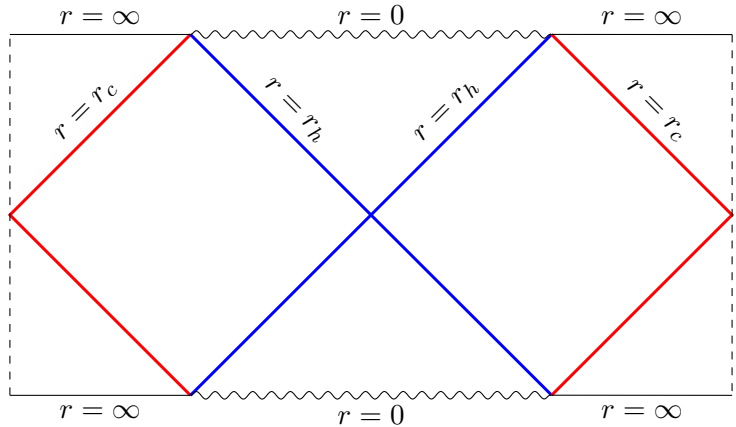
\begin{figure}[t!]
\centering
\begin{tikzpicture}[scale=1.2]
	\pgfmathsetmacro\myunit{4} 
           \draw [dashed, white]	(0,0)			coordinate (a)
		--++(90:\myunit)	coordinate (b);
	\draw [white] (b) --++(0:\myunit)		coordinate (c);
							
	\draw[dashed, white] (c) --++(-90:\myunit)	coordinate (d);
          \draw [line width = .4mm, red] (b)  --  node[pos=.5, above, sloped] {${\color{black} r=r_{c}}$} (-2,2) -- (a);		
	\draw [line width = .4mm, blue] (b)  --  node[pos=.5, above, sloped] {${\color{black} r=r_{h}}$} (2,2) -- (d);
           \draw [line width = .4mm, blue] (c) --  node[pos=.5, above, sloped] {${\color{black} r=r_{h}}$} (2,2) -- (a);
            \draw [line width = .4mm, red] (c) -- node[pos=.5, above, sloped] {${\color{black} r=r_{c}}$} (6,2) -- (d);
    \draw (a) -- (-2,0) coordinate (e)  node[pos=.5, below] {$r=\infty$}; 
    \draw[decorate, decoration={snake, amplitude=0.5mm, segment length=2.5mm}] (a) -- (d)   node[pos=.5, below] {$r=0$};
    \draw [dashed]  (e) -- (-2,4) coordinate (f);
    \draw (f) -- (b)   node[pos=.5, above] {$r=\infty$}; 
    \draw [decorate, decoration={snake, amplitude=0.5mm, segment length=2.5mm}] (b) -- (c)   node[pos=.5, above] {$r=0$};
    \draw  (c) -- (6,4) coordinate (g)   node[pos=.5, above] {$r=\infty$}; 
    \draw [dashed] (g) -- (6,0) coordinate (h);
    \draw (h) -- (d)  node[pos=.5, below] {$r=\infty$}; 
\end{tikzpicture}
\caption{\small Penrose diagram of a static, neutral quantum black hole in dS$_3$.}
\label{fig:pendS}
\end{figure}

\subsection{Nariai Limit}

When $\mu\neq0$, the metric function $H(r)$ (\ref{eq:Hrfunc}) takes the same form as the blackening factor of a four-dimensional Schwarzschild-de Sitter black hole \cite{articleKottler}. Thus,  we have a (smaller) black hole horizon $r=r_{h}$ aside from the cosmological horizon $r=r_{c}$, each as a positive root of $H(r)$, with $r_{h}<r_{c}$. Consequently, setting $H(r_{h})=H(r_{c})=0$, we may express the de Sitter radius $R_{3}$ and $\mu\ell$ entirely in terms of horizon radii $r_{h}$ and $r_{c}$,
\beq R_{3}^{2}=\frac{r_{c}^{3}-r_{h}^{3}}{r_{c}-r_{h}}=  r_c^2 +   r_h^2 +   r_c   r_h\;,\qquad \mu\ell=\frac{r_{h}r_{c}^{3}-r_{h}^{3}r_{c}}{r^{3}_{c}-r_{h}^{3}}=\frac{  r_c   r_h ( r_c +   r_h)}{   r_c^2 +  r_h^2 +   r_h  r_c}\;.\label{eq:R3muellrcrh}\eeq
Note the form factor $H(r)$ factorizes in terms of $r_{c}$ and $r_{h}$
\beq H(r)=\frac{1}{R_{3}^{2}r}(r R_{3}^{2}-r^{3}-\mu\ell R_{3}^{2})=\frac{1}{R_{3}^{2}r}(r-r_{h})(r_{c}-r)(r+r_{c}+r_{h})\;.\label{eq:Hrfactorized}\eeq
In the limit $r_{h}\to0$, we recover the pure $\text{dS}_{3}$ slicing (coinciding with $\mu=0$). Moreover, note that the curvature singularity at $r=0$ is hidden behind the induced black hole horizon (see the Penrose diagram in Fig.~\ref{fig:pendS}).

As $\mu\ell$ increases, the size of the black hole will increase until eventually it saturates the size of the cosmological horizon, the well-known Nariai limit \cite{Nariai99,Ginsparg:1982rs} of the SdS black hole, where $r_{c}=r_{h}\equiv r_{\text{N}}$, the Nariai radius. The resulting mass of the black hole forms an upper bound on $(\mu\ell)$, denoted $(\mu\ell)_{\text{N}}$, to avoid a naked singularity. By setting $H(r_{\text{N}})=H'(r_{\text{N}})=0$, we find the Nariai radius $r_{\text{N}}$ and maximum size of  $(\mu\ell)_{\text{N}}$
\beq r_{\text{N}}=\frac{1}{\sqrt{3}}R_{\text{3}}\;,\quad (\mu\ell)_{\text{N}}=\frac{2}{3}r_{\text{N}}=\frac{2}{3\sqrt{3}}R_{3}\;.\label{eq:rNdef}\eeq
Therefore, when $\ell\neq 0$ the Nariai limit places an upper bound on $\mu$.

\begin{figure}[t]
\begin{center}
\includegraphics[width=8.5cm]{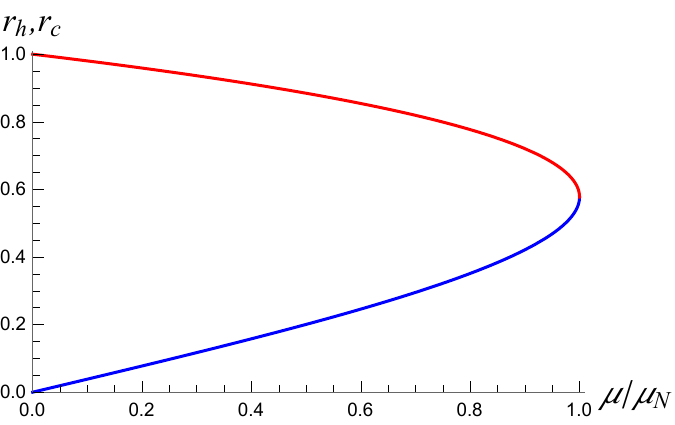}
\end{center}
\caption{\small Plot of the horizon radii $r_{h}$ (blue) and $r_{c}$ (red) as a function of $\mu/\mu_{\text{N}}$. The black hole horizon becomes larger as $\mu$ grows, while the cosmological horizon shrinks.}
\label{fig:rhrcvsmu} 
\end{figure}

Note that the relation (\ref{eq:R3muellrcrh}) may be inverted to find closed form expressions of $r_{h,c}$ in terms of $\mu$ and $r_{\text{N}}$ (see, \emph{e.g.}, \cite{Choudhury:2004ph,Morvan:2022ybp})
 \beq r_{h}=r_{\text{N}}(\cos\omega-\sqrt{3}\sin\omega)\;,\quad r_{c}=r_{\text{N}}(\cos\omega+\sqrt{3}\sin\omega)\;,\quad \text{with} \;\; \omega\equiv\frac{1}{3}\text{arccos}\left(\frac{\mu}{\mu_{\text{N}}}\right). \label{eq:rcrhinmu} \eeq
 This relation will prove  useful for plotting, see  Fig.~\ref{fig:rhrcvsmu}. Moreover, we see that the horizon radii are proportional to $r_{\text N}$ and hence to $\ell$, which is much larger than the Planck radius when $R_3 \gg \ell$, cf.~\eqref{eq:usefulapprox} (see also below).  

Thus, in the limit $\mu\ell\to(\mu\ell)_{\text{N}}$, the function $H(r)$ will have a double root at $r_{\text{N}}$. The region between these two roots describes the finite Nariai black hole solution.\footnote{See \cite{Horowitz:1996yb,Dias:2003up} for previous studies of the Nariai-like limit of the C-metric.} However, we cannot use coordinates $(t,r)$ to describe the Nariai metric since $H(r)$ vanishes in the region between the two horizons. To find the correct geometry, we follow \cite{Anninos:2012qw} and introduce dimensionful coordinates $(\tau,\rho)$ and a real, positive parameter $\beta$
\beq \tau=\epsilon t\;,\quad \rho=\frac{r-r_{h}}{\epsilon}\;,\quad \beta=\frac{r_{c}-r_{h}}{\epsilon}\;,\label{eq:taurhodef}\eeq
where $\epsilon$ is a dimensionful parameter quantifying the distance between radii $r_{c}$ and $r_{h}$. Taking the limit $r_{h}, r_{c}\to r_{\text{N}}$ and $\epsilon\to0$, one finds the Nariai  limit of the geometry
 (\ref{eq:AdS4Ccoord}),
\beq ds^{2}=\frac{\ell^{2}}{(\ell+xr_{\text{N}})^{2}}\left[-\frac{\rho(\beta-\rho)}{r_{\text{N}}^{2}}d\tau^{2}+r_{\text{N}}^{2}\frac{d\rho^{2}}{\rho(\beta-\rho)}+r_{\text{N}}^{2}\left(\frac{dx^{2}}{G(x)}+G(x)d\phi^{2}\right)\right]\;.\label{eq:AdS4CNariaiv1}\eeq
In these coordinates, the black hole horizon lives at $\rho=0$ while the cosmological horizon is at $\rho=\beta$. Further, performing the coordinate transformation 
\beq \tilde{\tau}=\frac{\beta}{2r_{\text{N}}}\tau\;,\quad \tilde{\rho}=\frac{2r_{\text{N}}}{\beta}(\rho-\beta/2)\;\label{eq:tautilderhotilde}\eeq
 brings the line element (\ref{eq:AdS4CNariaiv1}) to the form
\beq ds^{2}=\frac{\ell^{2}}{(\ell+xr_{\text{N}})^{2}}\left[-\left(1-\frac{\tilde{\rho}^{2}}{r_{\text{N}}^{2}}\right)d\tilde{\tau}^{2}+\left(1-\frac{\tilde{\rho}^{2}}{r_{\text{N}}^{2}}\right)^{-1}d\tilde{\rho}^{2}+r_{\text{N}}^{2}\left(\frac{dx^{2}}{G(x)}+G(x)d\phi^{2}\right)\right]\;.\label{eq:AdS4CNariaiv2}\eeq
The $(\tilde{\tau},\tilde{\rho})$ geometry  is two-dimensional de Sitter space with length scale $r_{\text{N}}$, while the $(x,\phi)$ sector   describes  (distorted) half-spheres of curvature radius $r_{\text{N}}$.


\subsection{Solution parameters and validity range: 4D and 3D views}

Let us pause for a moment to compare the two different ways of viewing the solution and the three parameters that characterize it. From the perspective of the bulk, these are naturally taken to be 
\begin{equation}
    \ell_4\,,\quad \ell\,,\quad \mu
\end{equation}
corresponding to the bulk cosmological radius, the brane tension, and the bulk black hole parameter. In the three-dimensional interpretation it is more natural to take them to be 
\begin{equation}
    R_3\,,\quad c G_3\,, \quad \mathcal{G}_3M
\end{equation}
namely, the dS$_3$ radius, the gravitational backreaction parameter, and the three-dimensional black hole mass. 
For convenience, we summarize here the relations between the two sets as
\begin{equation}
    \ell_4=2 c G_3 \,,\quad \textrm{or} \quad \ell=2 cG_3\left(1+\mathcal{O}(c G_3/R_3)^2\right)\,,
\end{equation}
\begin{equation}
    \frac1{\ell^2}-\frac1{\ell_4^2}=\frac1{R_3^2}\,,
\end{equation}
and
\begin{equation}
    \mu=1-\frac{1-x_1^2}{x_1^3}\,,\qquad 8\mathcal{G}_3M =1-\frac{4 x_1^2}{(3-x_1^2)^2}\,.
\end{equation}
The first relation is also given expanded for small backreaction $c G_3/R_3\ll 1$. The last one is in parametric form. The `renormalized' Newton's constant is
\begin{equation}
    \mathcal{G}_3=\frac{G_4}{2\ell}=\frac{\ell_4}{\ell} G_3 = G_3 + \mathcal{O}(c G_3/R_3)^2\,.
\end{equation}
Bear in mind, though, that the Newton constants, in three or four dimensions, do not enter as additional independent parameters of the solutions.

As we mentioned above, the effective theory expansion \eqref{indact} indicates that the cutoff length scale of the three-dimensional effective theory is $\ell_4$,  i.e., $c L_P$. When there is a large number of quantum  fields, $c\gg 1$, this length scale is much larger than the quantum gravity scale $L_P$. In other words, the cutoff energy is much lower than the naive one, a phenomenon already noted in \cite{Emparan:2002px} in the context of braneworld holography, and which is well known in more generality \cite{Dvali:2007hz}. The black holes that we have constructed have a size $\propto c L_P$, and their length scales never acquire values much larger than $c L_P$, e.g., $F(M)$ is never very large. Thus, in principle they may be subject to threshold effects of the effective theory.

On the other hand, from the bulk viewpoint there is no problem in considering black holes with sizes smaller than $\ell_4$ -- only that, since they involve sub-AdS scales, they are harder to understand from the dual CFT viewpoint. Our viewpoint in this article is that any AdS space can be viewed as a definition of a cutoff CFT. This does not guarantee the existence of its ultraviolet completion, but as long as the only degrees of freedom that are added at the cutoff scale are the KK graviton modes, then our black holes are accurately described by the bulk solution.\footnote{See \cite{Emparan:2002px} for a more extensive discussion.} The classical bulk description is reliable as long as the black holes are larger than the four-dimensional Planck length, which is the case when $c\gg 1$. 

 With these caveats in mind, we can regard our black holes as valid solutions of the quantum backreaction problem.

\subsection{Backreaction and quantum black holes}

With the metric (\ref{eq:qdS3}) in hand, we can, in principle, systematically compute the renormalized stress-tensor to all orders in $\ell$ using the right-hand side of the equations of motion (\ref{eq:semiclasseom}). To do this, we use  the expansion of length scales (\ref{eq:approx}) and perturbatively expand $\langle T^{\text{CFT}}_{\alpha\beta}\rangle$ in powers of $\ell^{2}$, such that
\beq \langle T^{\text{CFT}}_{\alpha\beta}\rangle=\langle T^{\text{CFT}}_{\alpha\beta}\rangle_{0}+\ell^{2}\langle T^{\text{CFT}}_{\alpha\beta}\rangle_{2}+...\;.\eeq
The analysis can be piggy-backed on the one in AdS$_3$ via  $\ell_{3}\to iR_{3}$, so we refer to \cite{Emparan:2020znc} for details.
%
The result is
\begin{align} \langle T^{\alpha}_{\;\beta}\rangle_{0}
=&\frac{c}{8\pi}\frac{ F(M)}{\bar{r}^{3}}\text{diag}(1,1,-2)
\;,\end{align}
where we have substituted $\ell=2cG_3$. This clearly has zero trace. With a little more work, the stress tensor at $\mathcal{O}(\ell^{2})$ is
\beq 
\begin{split}
\langle T^{\alpha}_{\;\beta}\rangle_{2}&=\frac{\ell}{16\pi G_{3}}\frac{F(M)}{\bar{r}^{3}}\biggr(-\frac{1}{2R_{3}^{2}}\text{diag}(1,-11,10)-\frac{24\mathcal{G}_{3}M}{\bar{r}^{2}}\text{diag}(3,1,-4)\\
&+\frac{\ell F(M)}{2\bar{r}^{3}}\text{diag}(-29,-17,43)\biggr)\;.
\end{split}
\label{eq:Talphabeta2}\eeq
This $\mathcal{O}(\ell^{2})$ contribution $\langle T^{\alpha}_{\;\beta}\rangle_{2}$ has a non-zero trace, thus breaking the conformal symmetry  due to the cutoff $\ell$. 

We could in principle continue this calculation and derive higher-order contributions to $\langle T_{\alpha\beta}^{\text{CFT}}\rangle$, simply by expanding the induced action to higher orders in~$\ell$. This is a distinct technical advantage over the more standard approach carried out in Sec.~\ref{sec:Qstresstensec}.
Comparing to \cite{Emparan:2020znc}, we see this is related to the backreaction on a quantum BTZ black hole background via the Wick rotation $\ell_{3}\to iR_{3}$.

 In the limit $\ell\to0$, the metric (\ref{eq:qdS3}) becomes the standard (no black hole) Schwarzschild-de Sitter solution in three dimensions. Thus, given the above discussion, for $\ell>0$, it is natural to interpret (\ref{eq:qdS3}) as a quantum black hole in three-dimensional de Sitter space, namely, the \emph{quantum Schwarzschild-de Sitter} solution (qSdS). 

 There are three special limits of the qSdS solution to consider. One is the Nariai limit, which we will describe in more detail momentarily. To describe the other two limits, note that $\mu$ and $\ell$ are technically independent parameters. This leads to the first important limit of the qSdS solution. Specifically, from the definition for $\mu$ (\ref{eq:defmu}), recall $\mu=0$ implies $x_{1}=1$. Equivalently, via (\ref{eq:Massx1}) and (\ref{eq:FMx1}), $x_{1}=1$ corresponds with $M=0$ and $F(M)=0$ on the brane, respectively. Hence $\mu=0$ removes the $O(1/r)$ term from the blackening factor along with the black hole horizon. Therefore, $\mu=0$ (or $x_{1}=1$) yields the three-dimensional \emph{quantum de Sitter} spacetime ($\text{qdS}_{3}$), accounting for the backreaction of quantum fields outside of the cosmological horizon when $\ell\neq0$. Lastly, in the infinite $R_{3}$ limit, the $O(r^{2}/R_{3}^{2})$ in the blackening factor is negligible, leading to a \emph{quantum Schwarzschild} black hole. In the next section we will be more careful with this limit, where we see small qSdS black holes (with $R_{3}\gg r_{h}$) behave  thermodynamically like a flat Schwarzschild solution.

\subsection*{Comparing $\langle T^{\alpha}_{\;\beta}\rangle$}

\begin{figure}[t!]
\begin{center}
\includegraphics[width=.42\textwidth]{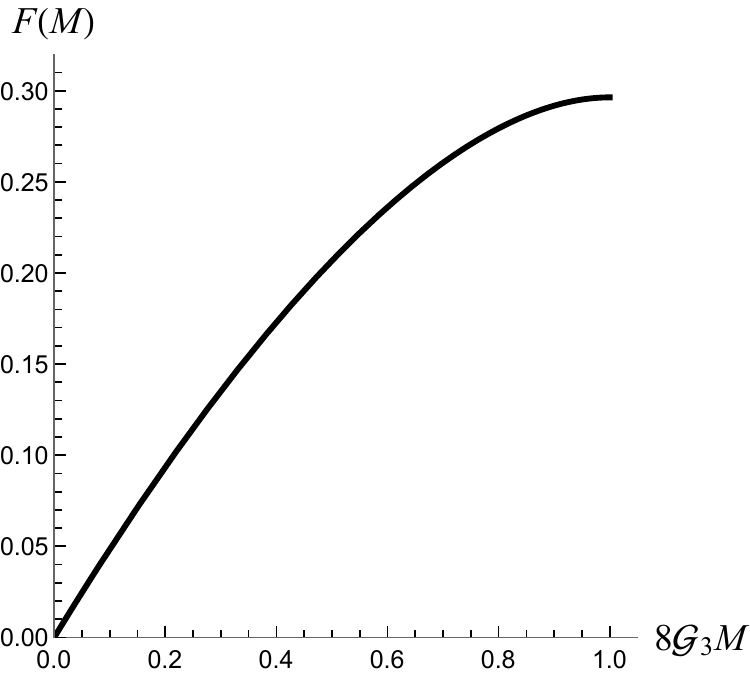}$\qquad\quad$\includegraphics[width=.42\textwidth]{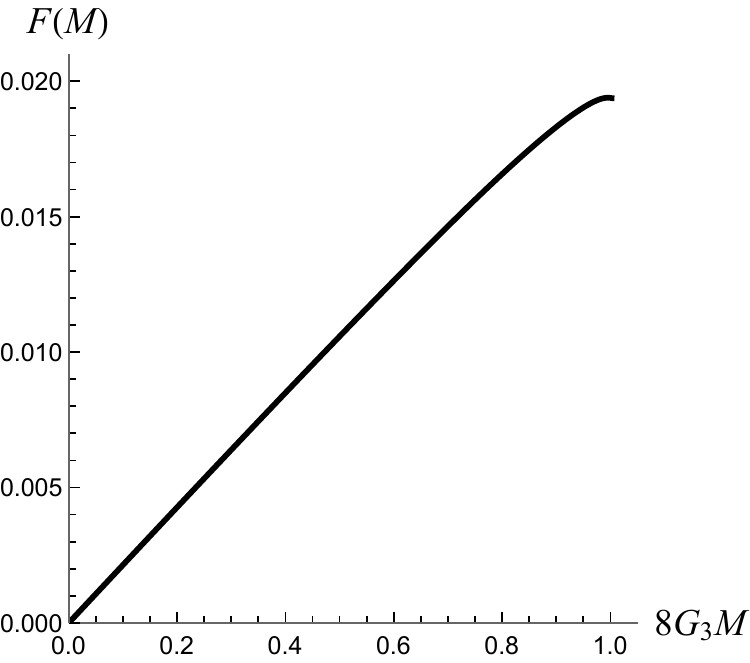}
\end{center}
\caption{\small \textbf{Left}: Form factor $F(M)$ computed holographically. When backreaction is present, the upper bound in the mass, set by the Nariai limit, is lower. \textbf{Right}: $F(M)$ found from renormalized stress tensor of a single conformal scalar in conical $\text{dS}_{3}$. For comparison, the holographic result must be multiplied by the central charge $c\gg 1$. }
\label{fig:FMfuncs} 
\end{figure}

The renormalized stress-energy tensor for the non-backreacted conformal fields turns out to have the form of \eqref{casstress}, whether we obtain it using holography or solving the quantum theory of a free conformal scalar in a conical geometry. Even though the two methods of calculation are completely different, they only differ in the mass dependence of the function $F(M)$, which is actually expected since each approach has different field content -- and actually, the shapes of $F(M)$ do not differ strongly, as we can see in Fig.~\ref{fig:FMfuncs}.

We can understand that in both cases $\langle T^{\alpha}_{\;\beta}\rangle \propto 1/r^3$. This behavior is natural given the scale invariance of the system, but it is not automatically dictated by it, since in a dS$_3$ geometry the stress tensor could, in addition to depending on $G_3 M$, also depend on $R_3/r$. However, as shown in \cite{Souradeep:1992ia}, the stress tensor for a conformal field in conical dS$_3$ can be obtained, by a Weyl transformation, from the stress tensor in conical Minkowski spacetime, and in the latter case, conformal symmetry does dictate the $1/r^3$ dependence. Observe that this radial dependence of the stress tensor gives rise to the $1/r$ corrections in the metric.

The stress tensors also have the same tensorial structure $\textrm{diag}(1,1,-2)$. Conformal invariance requires tracelessness, but this would still allow for another independent tensor structure, e.g., $\textrm{diag}(-2,1,1)$ (which would correspond to a thermal state). The agreement between our holographic and free field calculations simply reflects that in both cases the fields have transparent boundary conditions. For free fields, this was an explicit choice we made in \eqref{greens}, while these conditions are naturally selected in the holographic setup, since bulk fluctuations (dual to conformal field excitations) can freely travel in the bulk.

\subsection*{Nariai limit of the qSdS solution}

A chief difference between the quantum BTZ \cite{Emparan:2020znc} and qSdS solutions is that the latter has both a cosmological and black hole horizon. This is also in contrast to the classical three-dimensional  SdS solution, which only has a cosmological horizon.  Further, as we reviewed in the bulk geometry, there is a Nariai limit (\ref{eq:rNdef}) for which the black hole and cosmological horizons coincide, leading to the Nariai geometry of the $\text{AdS}_{4}$ C-metric (\ref{eq:AdS4CNariaiv2}). This geometry is also imprinted on the brane, such that the qSdS system has a well-defined Nariai limit. Following the same steps leading to (\ref{eq:AdS4CNariaiv2}), we can likewise find the Nariai limit of the quantum SdS black hole (\ref{eq:qdS3}). Specifically, one finds
\beq ds^{2}|_{x=0}=-\left(1-\frac{\tilde{\rho}^{2}}{\bar{r}_{\text{N}}^{2}}\right)d\tilde{\tau}^{2}+\left(1-\frac{\tilde{\rho}^{2}}{\bar{r}_{\text{N}}^{2}}\right)^{-1}d\tilde{\rho}^{2}+\bar{r}_{\text{N}}^{2}d\bar{\phi}^{2}\;,\label{eq:NariaiBHonbrane}\eeq
where $\tilde{\tau}=\frac{\bar{\beta}}{2\bar{r}_{\text{N}}}\tau$ and $\tilde{\rho}\to\eta\tilde{\rho}$. The geometry describes $\text{dS}_{2}\times S^{1}$, where the curvature radii of $\text{dS}_{2}$ and the circle $S^{1}$ are $\bar{r}_{\text{N}}$. 

As is well known for the classical four-dimensional SdS black hole, the Nariai limit of the qSdS solution yields an upper bound on the mass, denoted $M_{\text{N}}$. The mass can be determined as follows. First invert the definition of $\mu$ (\ref{eq:defmu}) to express $x_{1}(\mu)$. Upon substituting $x_{1}(\mu)$ into the definition of the mass (\ref{eq:Massx1}), we have $M(\mu)$. The Nariai mass is then $M_{\text{N}}=M(\mu_{\text{N}})$, where $\mu_{\text{N}}=\frac{2}{3\sqrt{3}}(R_{3}/\ell)$. While the resulting expression is complicated, it is a monotonically increasing function which is consistent with the range (\ref{eq:massrangev1}), \emph{i.e.}, $M_{\text{N}}<1/(8\mathcal{G}_{3})$, which comes from demanding that the conical deficit is less than $2\pi$. Precisely, for $R_{3}/\ell\to\infty$ (with $R_{3}>\ell$), then the function $M_{\text{N}}(R_{3}/\ell)$ is approximated by,
\beq 8 \mathcal G_3 M_{\text{N}} \approx 1 - \frac{2^{4/3}}{3} \frac{1}{(R_3 / \ell)^{2/3}} +...\;,\eeq
where the ellipsis corresponds to higher inverse powers of $(R_{3}/\ell)$. Moreover, in the limit $R_{3}\approx \ell$, we find a non-zero lower bound on the Nariai mass, 
\beq 8\mathcal{G}_{3}M_{\text{N}}\approx \frac{11}{27}+\frac{160}{729}(R_{3}/\ell-1)+...\;.\eeq
Combined, the Nariai mass lies in the range
\beq \frac{11}{27}<8\mathcal{G}_{3}M_{\text{N}}<1\;.\eeq
Thus, the Nariai bound $M < M_N$ is more stringent than the conical defect bound $M< 1/8 \mathcal G_3$ for finite $R_{3}/\ell$, coinciding only when $R_{3}/\ell\to\infty$. Consequently, due to gravitational quantum backreaction there is a limit on the amount of mass that one can put in dS$_3$, which does not saturate the maximum conical deficit angle. Since this is only due to the existence of a Nariai limit on the black hole solutions, it is likely to be present for non-holographic quantum black holes. However, the specific value of the mass bound depends on the details of the quantum theory.

Lastly, the Nariai solution places a bound on the quantum backreaction due to the CFT for which a quantum black hole in $\text{dS}_{3}$ exists. Particularly,  for non-vanishing backreaction there exists a maximum value of $F(M)$ (or equivalently $\mu$) via (\ref{eq:rNdef}), and thus a maximum value of $\Delta\phi$. If the deficit angle grows too large, then the backreaction creates a black hole too large to fit inside $\text{dS}_{3}$, such that the mass exceeds the Nariai bound. For such deficits, the quantum SdS solution no longer exists, but rather a naked conical defect spacetime with an \emph{unexcited} CFT. This means that in Fig.~\ref{fig:FMfuncs} the curve for $F(M)$ only extends in mass up to the Nariai bound, which depends on the backreaction parameter.


\section{Thermodynamics of the quantum SdS black hole} \label{sec:qSdSthermo}

Above we found a black hole solution on the brane $x=0$, which, from the brane perspective, is naturally interpreted as a quantum $\text{dS}_{3}$ black hole, whose black hole horizon is generated due to the presence of a backreacting CFT. The form of the metric is reminiscent of the classical four-dimensional SdS solution, owing to its origin as a black hole in a classical four-dimensional bulk.

Here we further our analysis of the qSdS solution by studying its horizon thermodynamics. In a certain respect, the qSdS thermodynamics is richer than the thermodynamics of the quantum BTZ solution since the qSdS solution has both a cosmological and black hole horizon. Consequently, we will find an entropy and temperature associated with each horizon, and explore the interplay between each system. The analysis below not only provides an important case study of the thermodynamics of quantum de Sitter black holes, it lends another non-trivial consistency check of braneworld holography.

\subsection*{Mass}


As discussed above, the system has three parameters, and once we fix a scale, there only remain two dimensionless parameters to characterize it. It is natural to take them as measuring the black hole size and the backreaction length scale $\ell$ in units of the dS$_3$ radius. 
Thus, we follow \cite{Emparan:1999fd,Emparan:2020znc} and introduce 
\beq z\equiv\frac{R_{3}}{r_{+}x_{1}}\;,\label{eq:zdef}\eeq
and 
\beq \nu\equiv\frac{\ell}{R_{3}}\;,\label{eq:nudef}\eeq
as convenient dimensionless parameters. The latter obviously measures the strength of backreaction for fixed $R_{3}$, while $z$, which is real and non-negative, is conveniently defined in terms of $r=r_{+}$, a positive real root of $H(r)$ for a horizon, and $x_1$, which characterizes the pole of the horizon along the bulk symmetry axis. We emphasize $r_{+}$ represents either the black hole horizon $r_{h}$ or the cosmological horizon $r_{c}$. Correspondingly, we often write $z_{h,c}=\frac{R_{3}}{r_{h,c}x_{1}}$.  

Two limits worth noting are: the limit of empty quantum de Sitter, which corresponds to $\mu=0$, i.e., $x_1=1$ and $r_+=R_3$, so we recover it for $z=1$; and the zero-backreaction limit, $\ell\to0$, in which there is no black hole, \emph{i.e.,} $r_{h}=0$, and $z_{h}\to+\infty$. 

We can now express $x_{1}$, $\mu$, and $r_{+}$ solely in terms of parameters (\ref{eq:zdef}) and (\ref{eq:nudef}). 
Solving $H(r_{+})=0$ for $x_{1}^{2}$   yields
\beq x_{1}^{2}=\frac{1}{z^{2}}\frac{1+\nu z^{3}}{1+\nu z}\;.\label{eq:x1sqrel}\eeq
Rearranging $z$ (\ref{eq:zdef}), squaring and substituting in $x_{1}^{2}$ (\ref{eq:x1sqrel}) leads to \beq r_{+}^{2}=R_{3}^{2}\frac{1+\nu z}{1+\nu z^{3}}\;.\label{eq:rpdef}\eeq
Similarly,  from $\mu=(1-x_{1}^{2})/x_{1}^{3}$, 
we find 
\beq \mu x_{1}=\frac{z^{2}-1}{1+\nu z^{3}}\;.\label{eq:mux1}\eeq
Notice the quantities (\ref{eq:x1sqrel}), (\ref{eq:rpdef}), and (\ref{eq:mux1}) are the Wick rotated counterparts of the qBTZ solution \cite{Emparan:2020znc}.\footnote{Specifically, with $\ell_{3}^{2}\to -R_{3}^{2}$, we have $z^{2}\to -z^{2}$, $\nu^{2}\to-\nu^{2}$ and $\nu z\to \nu z$.} Lastly, in terms of $\nu$ and $z$, the relation between the bare Newton's constants $G_{4}$ and $G_{3}$ is
\beq G_{4}=2\ell_{4}G_{3} 
=\frac{2G_{3}\ell}{\sqrt{1-\nu^{2}}}\;,\label{eq:G4G3rel}\eeq
while the renormalized Newton's constant (\ref{eq:Massx1}) is 
\beq \mathcal{G}_{3}=\frac{\ell_{4}}{\ell}G_{3}=\frac{G_{3}}{\sqrt{1-\nu^{2}}}\;.\eeq
Putting together the relations (\ref{eq:x1sqrel}), (\ref{eq:rpdef}), (\ref{eq:mux1}), the mass $M$  (\ref{eq:Massx1})  is recast as
\beq 
M 
=\frac{1}{8G_{3}}\sqrt{1-\nu^{2}}\frac{(z^{2}-1)(9z^{2}-1+8\nu z^{3})}{(3z^{2}-1+2\nu z^{3})^{2}}\;.
\label{eq:mass3}\eeq
 Further, the function $F(M)$ \eqref{eq:FMx1} in terms of $z$ and $\nu$ is
\beq
F(M)= \frac{8 z^4 (z^2-1)(1 + \nu z)^2}{(3z^2 - 1 + 2 \nu z^3)^3}\;.
\eeq


In the quantum de Sitter limit, where $z=1$, 
we have $M=0$ and $F(M)=0$, as expected.
It is also worth pointing out that the mass $M$ will vanish at large $z$, \beq \lim_{z\to\infty} M\approx \frac{1}{4\mathcal{G}_{3}\nu z}+\mathcal{O}(1/z^{2})\;.\eeq
For fixed $x_{1}\neq 1$ and for $z=z_{h}$, it is natural to think of the large $z_{h}$ limit as a \emph{small} quantum Schwarzschild black hole, where $R_{3}\gg r_{h}  x_{1}$. We will revisit this special case momentarily.

There are two additional limits of the mass (\ref{eq:mass3}) worth emphasizing. First, in the limit of vanishing backreaction, $\nu\to0$, the mass $M$ simplifies to 
\beq \lim_{\nu\to0}M=\frac{1}{8G_{3}}\frac{(z^{2}-1)(9z^{2}-1)}{(3z^{2}-1)^{2}}\;.\label{eq:Mnu0}\eeq
We will utilize this relation below shortly. Second, recall that the black hole has a Nariai limit (\ref{eq:NariaiBHonbrane}), where $(\mu\ell)$ attains a maximum (\ref{eq:rNdef}), 
such that 
\beq \mu_{\text{N}}=\frac{2}{3\sqrt{3}}\frac{1}{\nu}\;.\label{eq:muNnu}\eeq
Since in the Nariai limit, $r_{h}=r_{c}=r_{\text{N}}$, we can introduce a $z_{\text{N}}$, 
\beq z_{\text{N}}=\frac{R_{3}}{r_{\text{N}}x_{1}^{\text{N}}}=\frac{\sqrt{3}}{x_{1}^{\text{N}}}\;,\eeq
where $x_{1}^{\text{N}}$ is the particular value of $x_{1}$ in the Nariai limit, found by   solving $\mu_{\text{N}}=(1-x_{1}^{2})/x_{1}^{3}$ for $x_{1}$. For arbitrary $\nu$, there will generally be two complex solutions of $x_{1}$ and one real solution, which for small $\nu$ takes the form
\beq x_{1}^{\text{N}}=\sqrt{3}\left(\frac{\nu}{2}\right)^{1/3}-\frac{\sqrt{3}}{2}\nu+\mathcal{O}(\nu^{5/3})\;,\eeq
and vanishes in the limit $\nu\to0$. The mass of the Nariai solution $M_{\text{N}}$ is therefore defined by $M_{\text{N}}\equiv M|_{z=z_{\text{N}}}$. Further, in the limit $\nu\to0$, we find $M_{\text{N}}=\frac{1}{8G_{3}}$, which is equivalent to the large-$z$ limit of (\ref{eq:Mnu0}).

\subsection*{Temperature}

From the brane perspective, the black hole and cosmological horizons will appear to emit radiation at the Hawking and Gibbons-Hawking temperatures $T_{h},T_{c}$, respectively, 
\beq T_{h,c}=\frac{\kappa_{h,c}}{2\pi}\;,\eeq
where $\kappa_{h,c}$ are the associated surface gravities, defined by $\xi^{\mu}\nabla_{\mu}\xi^{\nu}=\kappa\xi^{\nu}$, where $\xi$ is the time-translation Killing vector, $\xi=\partial_{\bar{t}}$. Thus, in terms of $\nu$ and $z$, 
\beq 
\begin{split}
T_{h,c}&=\frac{|H'(\bar{r}_{+})|}{4\pi}=\frac{1}{4\pi}\biggr|\frac{\mu\ell\eta^{3}}{\bar{r}_{+}^{2}}-\frac{2\bar{r}_{+}}{R_{3}^{2}}\biggr|=\frac{z}{2\pi R_{3}}\frac{|2+3\nu z-\nu z^{3}|}{3z^{2}-1+2\nu z^{3}}\;.
\end{split}
\label{eq:tempT}\eeq
More explicitly, the temperatures $T_{h}$ and $T_{c}$ are given by
\beq 
T_{h } = -\frac{z_h}{2\pi R_{3}}\frac{ 2+3\nu z_{h}-\nu z_{h}^{3} }{3z_{h}^{2}-1+2\nu z_{h}^{3}}\;, \qquad T_{c } =\frac{z_c}{2\pi R_{3}}\frac{ 2+3\nu z_{c}-\nu z_{c}^{3} }{3z_{c}^{2}-1+2\nu z_{c}^{3}}\,.\eeq
 In the limit $\nu\to0$,   the black hole   temperature vanishes, since $z_h$ blows up in that case.  The cosmological horizon temperature reduce in that limit to 
\beq \lim_{\nu\to0}T_{ c}= \frac{1}{\pi R_{3}}\frac{z_{c}}{3z_{c}^{2}-1}\;.\label{eq:Tnu0lim}\eeq
Further, for non-zero $\nu\neq0$, we see the cosmological horizon of the quantum de Sitter solution is simply the Gibbons-Hawking temperature, $T_{c}|_{z=1}=1/(2\pi R_{3})$, \emph{i.e.}, the backreaction does not alter the temperature in $\text{qdS}_{3}$.


Since $r_{h}<r_{c}$, we have $z_{c}<z_{h}$, and, consequently, the black hole horizon is hotter than the cosmological horizon $T_{h}>T_{c}$. Thus, as usual for Schwarzschild-de Sitter spacetimes, the black hole and cosmological horizons are not in thermal equilibrium. Only in the Nariai limit do the temperatures agree with each other, where the system is in thermal equilibrium.

The Nariai limit, however,  is subtle because the surface gravities associated with the Killing vector $\partial_{\bar t}$ vanish in this limit. To see this, note that the surface gravities $\kappa_{h}$ and $\kappa_{c}$ for the normalization $\xi=\partial_{\bar{t}}$ are 
\beq 
\begin{split}
&\kappa_{h}=\frac{1}{2}H'(\bar{r}_{h})=\frac{1}{2\bar{r}_{h}r_{\text{N}}^{2}}(\bar{r}_{\text{N}}^{2}-\bar{r}_{h}^{2})\;,\quad \kappa_{c}=-\frac{1}{2}H'(\bar{r}_{c})=-\frac{1}{2\bar{r}_{c}r_{\text{N}}^{2}}(\bar{r}_{\text{N}}^{2}-\bar{r}_{c}^{2})\;,
\end{split}
\eeq
which vanish in the limit $\bar{r}_{h,c}\to \bar{r}_{\text{N}}$. Consequently, $T_{h,c}\to0$ as $\bar{r}_{h,c}\to \bar{r}_{\text{N}}$. If, however, the Killing vector is normalized as $\xi^2 =-1$ at the radius $\bar r_0$ where the blackening factor $H(\bar r)$ obtains a maximum, then the surface gravities are non-vanishing in the Nariai limit \cite{Bousso:1996au}. Specifically,
\beq H'(\bar{r}_{0})=0 \quad \rightarrow \quad  \bar r^3_0= \frac{1}{2} \ell F(M) R_{3}^{2}=\frac{\bar{r}_{+}}{2}(3\bar{r}_{\text{N}}^{2}-\bar{r}_{+}^{2}) \;.\eeq
Then,  
\beq
\begin{split}
H(\bar r_0) &= 1-8\mathcal{G}_{3}M-\frac{\ell F(M)}{\bar{r_0}}-\frac{\bar{r_0}^{2}}{R_{3}^{2}} 
=\frac{1}{r_{\text{N}}^{2}}(\bar{r}_{\text{N}}^{2}-\bar{r}_{0}^{2})\;. 
\end{split}
\eeq
The new temperature $\bar{T}$, which was introduced in \cite{Bousso:1996au}, is
\beq
\bar{T}= \frac{T}{\sqrt{H(\bar r_0)}}\;.\eeq
In terms of the horizon radii $\bar{r}_{h,c}$, we have
\beq 
\begin{split}
&\bar{T}_{h,c}
=\mp\frac{1}{4\pi r_{\text{N}}\bar{r}_{h,c}}\frac{\bar{r}_{\text{N}}^{2}-\bar{r}_{h}^{2}}{\sqrt{\bar{r}_{\text{N}}^{2}-\left(\frac{\bar{r}_{h,c}}{2}(3\bar{r}^{2}_{\text{N}}-\bar{r}_{h,c}^{2})\right)^{2/3}}}\;, 
\end{split}
\eeq
 \begin{figure}[t!]
\begin{center}
\includegraphics[width=7.5cm]{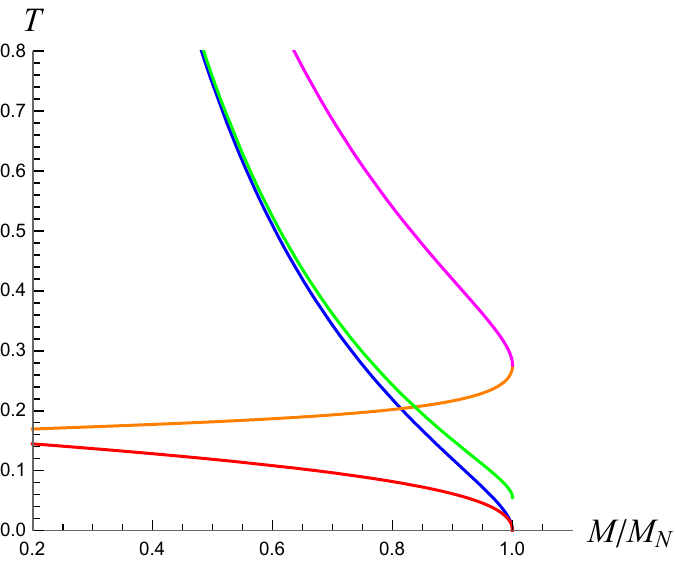}
\end{center}
\caption{\small Plot of the temperature as a function of mass $M$ for $\nu=1/3$. The blue and red curves correspond to temperatures $T_{h}$ and $T_{c}$, respectively, while the magenta and orange curves correspond to $\bar{T}_{h}$ and $\bar{T}_{c}$, respectively. The green curve represents the temperature of the Schwarzschild limit, where $R_{3}\gg r_{h}$, coinciding with $T_{h}$ for small mass qSdS black holes.}
\label{fig:TvsM} 
\end{figure}
where the minus sign corresponds to the black hole temperature, and the plus sign to the cosmological horizon temperature.
Carefully taking the limit $\bar{r}_{\text{N}}\approx \bar{r}_{h,c}$,  the temperature $\bar{T}_{h,c}$ of both horizons approaches the Nariai temperature $T_{\text{N}}=1/(2\pi r_{\text{N}})$ (see App.~B in \cite{Svesko:2022txo}). The expression for $\bar{T}$ in terms of $z$ is cumbersome, however, in the small $\nu$ limit we have
 \beq \bar{T}|_{\nu\approx0}=\frac{1}{2\pi R_{3}}\left[1+\frac{3}{2}\left(\frac{z(z^{2}-1)\nu}{2}\right)^{2/3}+\mathcal{O}(\nu)+...\right]\;,\eeq
 while for $z=1$, $\bar{T}=1/(2\pi R_{3})$, the Gibbons-Hawking temperature of $\text{dS}_{3}$, as expected.

In Fig.~\ref{fig:TvsM} we plot the temperatures $T_{c},T_{h}$, and $\bar{T}$  as a function of mass $M$. The behavior of the temperatures are essentially identical to that of classical four-dimensional Schwarzschild-de Sitter black holes (see, for instance Fig.~2 of \cite{Morvan:2022ybp}). In particular, notice for small mass $M$ black holes, the black hole temperature diverges, such that it may be approximated by the temperature of a (quantum) Schwarzschild black hole,  
\beq T_{\text{Schwarz}}=\frac{1}{2\pi R_{3}}\frac{\nu z_{h}^{2}(z_{h}^{2}-1)}{(-1+3z_{h}^{2}+2\nu z_{h}^{3})}\;,\label{eq:Schwarztemp}\eeq
which follows from dropping the second term in the first equality of (\ref{eq:tempT}), using the approximation $R_{3}\gg r_{h}$. As seen in Fig.~\ref{fig:TvsM}, the black hole temperature $T_{h}$ of the qSdS solution approaches the Schwarzschild temperature (\ref{eq:Schwarztemp}) at small mass. Collectively, our observations indicate the three-dimensional quantum Schwarzschild-de Sitter system, at least thermodynamically, behaves like a four-dimensional classical SdS spacetime, reflecting the holographic character of our setup. We will see additional evidence of this below.

\subsection*{Entropy}

For $\mu\neq0$, the bulk will have two horizons, $r_{h}$ and $r_{c}$.  The entropy $S_{\text{BH}}^{(4)}(r_{+})$ of either bulk horizon is given by the (four-dimensional) Bekenstein-Hawking entropy-area relation
\beq
\begin{split}
S_{\text{BH}}^{(4)}&=\frac{\text{Area}(r_{+})}{4G_{4}}=\frac{2}{4G_{4}}\int_{0}^{2\pi\eta}d\phi\int_{0}^{x_{1}}dx r_{+}^{2}\frac{\ell^{2}}{(\ell+xr_{+})^{2}}\\
&=\frac{\pi R_{3}}{G_{3}}\frac{z\sqrt{1-\nu^{2}}}{3z^{2}-1+2\nu z^{3}}\;,
\end{split}
\label{eq:Sgen}\eeq
where in the final line we used that $G_{4}$ is related to $G_{3}$ via (\ref{eq:G4G3rel}).
Again, here it is understood $z$ represents either $z_{h}$ or $z_{c}$, such that the entropy is localized around each horizon. 

From the brane perspective, this quantity is interpreted as the sum of the gravitational entropy plus the entanglement entropy due to the backreacting CFT, \emph{i.e.}, the three-dimensional generalized entropy $S_{\text{gen}}^{(3)}$,
\beq S_{\text{BH}}^{(4)}=S_{\text{gen}}^{(3)}\;.\label{eq:SBH4Sgen3}\eeq
Note that $S_{\text{gen}}^{(3)}$ is valid to all orders in $\nu$ since it is defined as a bulk magnitude, which is \emph{exact}, up to bulk quantum corrections. In the limit $z=1$, we have the generalized entropy of the quantum de Sitter solution, 
\beq S_{\text{gen}}^{(3)}\big|_{z=1}=\frac{2\pi R_{3}}{4G_{3}}\frac{\sqrt{1-\nu^{2}}}{1+\nu}\;,\label{eq:SgenqdS}\eeq
which is the same as \eqref{compacthor}, and is proportional to the Gibbons-Hawking entropy of the $\text{dS}_{3}$ cosmological horizon. As a generalized entropy, this quantity represents the sum of gravitational entropy and entanglement entropy due to the CFT living outside of the cosmological horizon. 

\begin{figure}[t]
\begin{center}
\includegraphics[width=5.5cm]{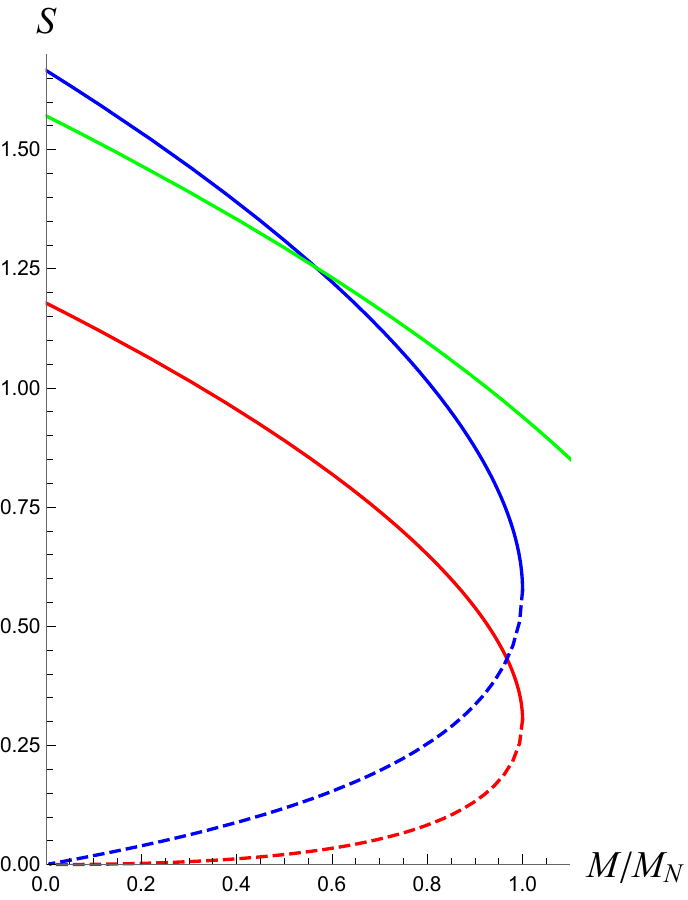}$\qquad\qquad\quad$\includegraphics[width=5.5cm]{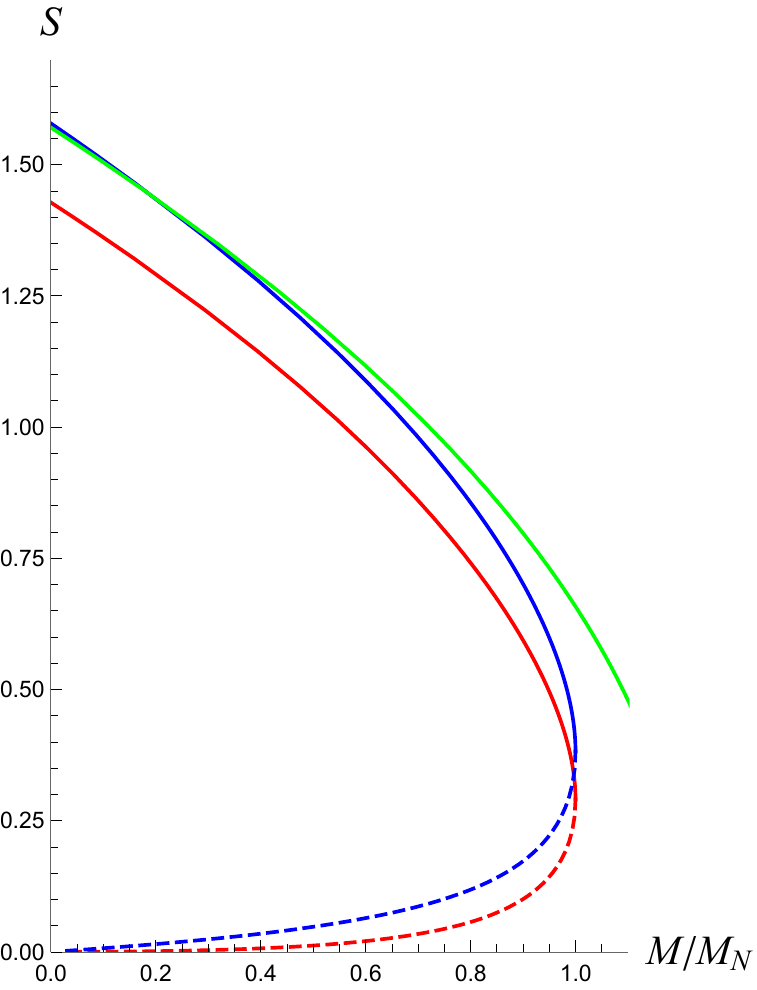}
\end{center}
\caption{\small Plot of $S_{\text{gen}}^{(3)}$ (red), $S_{\text{BH}}^{(3)}$ (blue) and $S_{\text{SdS}_{3}}$ (green) as a function of mass $M$. The dashed curves refer to black hole entropies $S_{h}$, while the solid curves denote the entropies associated with the cosmological horizon $S_{c}$. The backreaction parameter is $\nu=1/3$ (left) and $\nu=1/10$ (right), and $\mathcal{G}_{3}=R_{3}=1$. As $\nu\to0$, all solid curves collapse to $S_{\text{SdS}_{3}}$, while the dashed curves go to zero. 
}
\label{fig:SvsM} 
\end{figure}

\begin{figure}[t]
\begin{center}
\includegraphics[width=6cm]{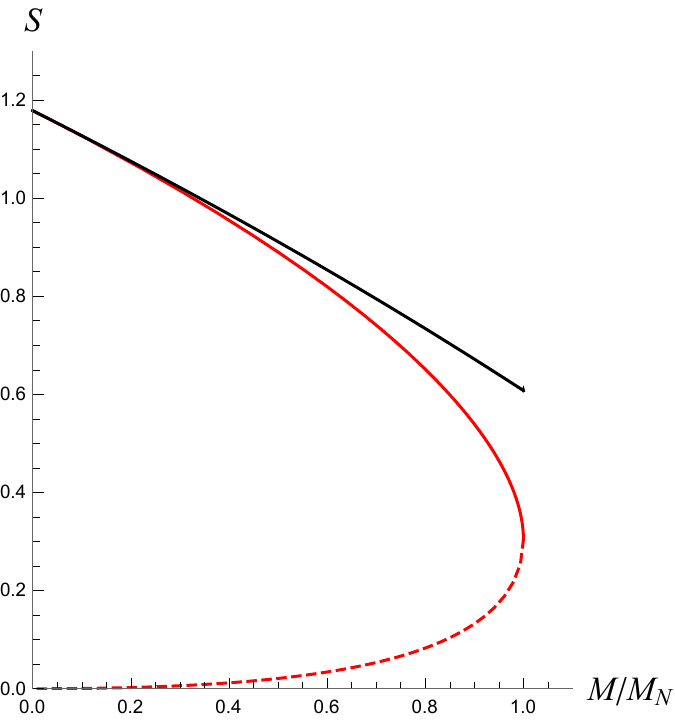}
\end{center}
\caption{\small Plot of generalized entropies $S_{\text{gen},h}^{(3)}$ (red dashed curve) and $S_{\text{gen},c}^{(3)}$ (red solid curve) as a function of $\mathcal{M}=M/M_{\text{N}}$. Notice the total generalized entropy $S_{\text{gen},\text{tot}}^{(3)}$ (black), given by the sum of the red solid and dashed curves, is approximately   a linear function of the mass. Here we have set $\nu=1/3$.}
\label{fig:Sgenwithtot} 
\end{figure}

Further, the generalized entropy (\ref{eq:SBH4Sgen3}) is related to the three-dimensional Bekenstein-Hawking entropy $S_{\text{BH}}^{(3)}$ of the horizon(s) on the brane as 
\beq
S_{\text{gen}}^{(3)} = \frac{\sqrt{1- \nu^2}}{1 + \nu z}  S_{\text{BH}}^{(3)}\;,
\label{eq:SgenasScl}\eeq
where $S_{\text{BH}}^{(3)}=\frac{2\pi r_{+}\eta}{4G_{3}}$. We see $S_{\text{BH}}^{(3)}$ is influenced by the backreaction, thus containing semi-classical quantum effects. In the limit of vanishing backreaction, the black hole horizon disappears and for the cosmological horizon we find $S_{\text{gen},c}^{(3)}$ and $S_{\text{BH},c}^{(3)}$ coincide and are proportional to the temperature $T_c$ (\ref{eq:Tnu0lim}) 
\beq \lim_{\nu\to0}S_{\text{gen},c}^{(3)}=\lim_{\nu\to0}S_{\text{BH},c}^{(3)}=\frac{\pi R_{3}}{G_{3}}\frac{z_c}{3z_c^{2}-1}=\frac{\pi^{2}R_{3}^{2}}{G_{3}}\lim_{\nu\to0} T_c=\frac{\pi R_{3}}{2G_{3}}\sqrt{1-8G_{3}M}\;,\eeq
where to arrive at the last equality we used that the temperature of the cosmological horizon with conical deficit is $T_c|_{\nu=0}=\frac{1}{2\pi R_{3}}\sqrt{1-8G_{3}M}$.
We recognize the entropy in this limit as the entropy of the classical three-dimensional Schwarzschild-de Sitter solution \cite{Spradlin:2001pw}
\beq S_{\text{SdS}_{3}}=\frac{2\pi r^{\text{SdS}_{3}}_{c}}{4G_{3}}\;,\eeq
with $r_{c}^{\text{SdS}_{3}}=R_{3}\sqrt{1-8G_{3}M}$. Finally, in the quantum de Sitter limit ($z=1$), the area entropy is simply equal to the Gibbons-Hawking entropy
\beq S_{\text{BH},c}^{(3)}|_{z=1}=\frac{2\pi R_{3}}{4G_{3}}=\frac{2\pi R_{3}}{4\mathcal{G}_{3}}\frac{1}{\sqrt{1-\nu^{2}}}\;,\eeq
where $\mathcal{G}_{3}$ is the renormalized Newton's constant. Therefore, the Gibbons-Hawking entropy of $\text{qdS}_{3}$ scales like the classical entropy of $\text{dS}_{3}$. 
A plot of each of these entropies $S_{\text{gen}}^{(3)},S_{\text{BH}}^{(3)}$, and $S_{\text{SdS}_{3}}$ is given in Fig.~\ref{fig:SvsM}. We observe that the sum of the black hole and cosmological horizon entropies $S_{\text{gen},h}^{(3)}$ and $S_{\text{gen},c}^{(3)}$ produces an approximately linear curve always equal to or less than the entropy of the quantum de Sitter solution (\ref{eq:SgenqdS}), see Fig.~\ref{fig:Sgenwithtot}. This is reminiscent of the observation  in    \cite{Visser:2019muv,Morvan:2022ybp} for the classical SdS solution that the   sum of the horizon entropies is approximately a linear function of the mass.   We will return to this point in Sec.~\ref{sec:entdeficit}, as it will prove useful when computing the nucleation rate of quantum dS black holes.

We can read off the leading order effect the CFT has on the Bekenstein-Hawking entropy by computing the difference between $S_{\text{BH}}^{(3)}$ and $S_{\text{SdS}_{3}}$,
\beq S_{\text{BH}}^{(3)}-S_{\text{SdS}_{3}}\approx \frac{\nu z(z^{2}-1)}{(3z^{2}-1)}S_{\text{SdS}_{3}}+\mathcal{O}(\nu^{2})\;.\label{eq:entdiff}\eeq
Recall from the relation between the central charge $c$ and $\nu$ (\ref{eq:usefulapprox}), that $c\approx \frac{\nu R_{3}}{2G_{3}}$, and thus the difference is linear in $c$. It is thus natural to interpret this difference as the leading contribution to the entanglement entropy of the CFT. We will see more evidence of this momentarily.

In addition to the leading order effect in $\nu$, the backreaction of the CFT induces higher curvature corrections, which enter at order $\mathcal{O}(\nu^{2})$. From the brane perspective, one computes the gravitational entropy due to this higher curvature corrections using Wald's entropy functional $S_{\text{Wald}}$ \cite{Wald:1993nt}
\beq S_{\text{Wald}}=-2\pi\int_{\mathcal{H}}dA\frac{\partial\mathcal{L}}{\partial R^{abcd}}\epsilon_{ab}\epsilon_{cd}\;,\eeq
where $dA=d^{d-2}x\sqrt{q}$ is the area element of a cross-section $\mathcal{H}$ of the horizon, with $q_{ab}$ being the induced metric on the cross-section,   $\mathcal{L}$ is the Lagrangian density defining the gravitational theory, and $\epsilon_{ab}$ is the binormal to $\mathcal H$. 
With respect to the induced gravity action (\ref{eq:totalact}), the gravitational entropy is \cite{Jacobson:1993vj}
\beq S_{\text{Wald}}^{(3)}=\frac{1}{4G_{3}}\int dx\sqrt{q}\left[1+\ell^{2}\left(\frac{3}{4}\tilde{R}-g^{ab}_{\perp}\tilde{R}_{ab}\right)+\mathcal{O}(\ell^{4}/R^{6}_{3})\right]\;.\label{eq:SWald}\eeq
 with $g^{\perp}_{bd}=g_{bd}-q_{bd}$ being the   metric in the directions orthogonal to the horizon.
 The dominant contribution is the three-dimensional Bekenstein-Hawking entropy
\beq S_{\text{Wald}}^{(3)}=\frac{1}{4G_{3}}\int dx\sqrt{q}=\frac{1}{4G_{3}}\int_{0}^{2\pi}d\bar{\phi}\bar{r}_{+}=S_{\text{BH}}^{(3)}\;,\eeq
as expected. Evaluating (\ref{eq:SWald}) on the qSdS background ($\bar{t},\bar{r},\bar{\phi})$, yields
\beq 
\begin{split}
S_{\text{Wald}}^{(3)} &=
\left[1+\frac{\nu^{2}}{2}-\nu^{3}\frac{z(z^{2}-1)}{(1+\nu z)}+\mathcal{O}(\nu^{4})\right]S_{\text{BH}}^{(3)}\;.
\end{split}
\label{eq:Waldentonshell}\eeq
 Clearly, in the limit of no backreaction, $S_{\text{Wald}}^{(3)}$ reduces to $S_{\text{BH}}^{(3)}$. Also notice the leading order higher curvature contributions are of order $\mathcal{O}(\nu^{2})$, in contrast to the linear order effect of the CFT in (\ref{eq:entdiff}).


It is well known that the generalized entropy is equal to the sum of the gravitational (Wald) entropy plus the fine-grained entropy of matter $S_{\text{out}}$ outside of the horizon. We can compute the three-dimensional matter entropy $S_{\text{out}}^{(3)}$ by taking the difference of $S_{\text{gen}}^{(3)}$ (\ref{eq:SgenasScl}) and the Wald entropy (\ref{eq:Waldentonshell}). To leading order in $\nu$ we have, 
\beq S_{\text{out}}^{(3)}=S_{\text{gen}}^{(3)}-S_{\text{Wald}}^{(3)}\approx-\nu zS_{\text{BH}}^{(3)}\;.\eeq
Since this quantity is linear in $\nu$, we see it is proportional to central charge $c$.
Notice for large $z$ but fixed $\nu z\ll1$ we find
\beq S_{\text{out}}^{(3)}\approx -\frac{2\pi c}{3}\;,\eeq
identical to what was found for the qBTZ geometry \cite{Emparan:2020znc}. As in that case, the overall minus sign does not imply the von Neumann entropy of the CFT is negative. Rather, $S_{\text{out}}^{(3)}$ only corresponds to the finite contribution to the CFT entropy upon absorbing the leading term in the renormalization of $G_{3}$.


\subsection*{First law}

Putting together the mass (\ref{eq:mass3}), temperature (\ref{eq:tempT}) and entropy (\ref{eq:Sgen}) for the black hole horizon, we find
\beq \partial_{z}M=T_h\partial_{z}S^{(3)}_{\text{gen},h}\;.\eeq
Keeping all other parameters fixed, we thus have the first law of (semi-classical) black hole thermodynamics
\beq dM=T_hdS^{(3)}_{\text{gen},h}\;.\label{eq:bhfirstlaw}\eeq
Similarly, for the cosmological horizon we have
\beq
dM = - T_c dS^{(3)}_{\text{gen},c}\;.
\label{eq:cfirstlaw}\eeq
Combining these variational relations we attain
\beq
0=T_h dS^{(3)}_{\text{gen},h} + T_c d S^{(3)}_{\text{gen},c}\;.
\label{eq:firstlawsum}\eeq
Thus, the thermodynamic quantities for either the black hole or cosmological horizon are not independent. Specifically, as the generalized entropy associated with the black hole increases, the generalized entropy attributed to the cosmological horizon decreases. Further, the minus sign  in the first law for the cosmological horizon (\ref{eq:cfirstlaw}) indicates the entropy of the cosmological horizon decreases as the mass increases. Consequently, the (generalized) entropy of quantum $\text{dS}_{3}$ is a maximum entropy configuration such that $\text{qdS}_{3}$ is an equilibrium state with a \emph{finite} number of degrees of freedom.

Each of the first laws (\ref{eq:bhfirstlaw}), (\ref{eq:cfirstlaw}), and (\ref{eq:firstlawsum})  are precisely what happens for classical higher dimensional SdS, except here the classical entropies $S_{h,c}$ have been replaced by their generalized counterparts $S_{\text{gen},h,c}$. The semi-classical first laws above also hold in the two-dimensional context, where one considers  de Sitter JT gravity \cite{Svesko:2022txo}. Consistent with the thermodynamics of the quantum BTZ solution \cite{Emparan:2020znc}, our observations here provide further evidence that, when semi-classical backreaction is accounted for, quantum black holes obey a first law of thermodynamics where the classical entropy is replaced by the generalized entropy. Additionally, we see the usual but peculiar minus sign in front of the cosmological first law (\ref{eq:cfirstlaw}) is present even when backreaction is accounted for, implying the thermodynamic interpretation of the minus sign is not resolved due to semi-classical modifications. 

\section{Entropy deficit and nucleation rate of quantum dS black holes}
\label{sec:entdeficit}

As with black holes, one expects the thermodynamics of the dS cosmological horizon to have a microscopic interpretation. A complete understanding of de Sitter geometry is a task for full-fledged quantum gravity, however, a promising explanation is offered by holography of the dS static patch \cite{Banks:2005bm,Parikh:2004wh,Anninos:2011af,Anninos:2017hhn,Leuven:2018ejp,Coleman:2021nor,Susskind:2021omt}. Recently it was proposed that the dual microscopic theory lives on the (stretched) cosmological horizon \cite{Susskind:2021omt,Susskind:2021dfc, Shaghoulian:2021cef}.\footnote{There is  another picture of static patch holography, where the dual quantum theory lives on a holographic screen near the north or south poles of the static patch \cite{Anninos:2011af,Anninos:2017hhn}. The two proposals are consistent if the cosmological horizon represents the IR of the underlying microscopic theory while the screen near the poles represents the UV \cite{Leuven:2018ejp}.} Evidence for this comes from studying, in particular, the entropy deficit generated by nucleating a four-dimensional classical black hole in de Sitter space, where the nucleation rate is controlled by the deficit. As we will review, the form of the entropy deficit suggests dS gravity has a matrix theory interpretation \cite{Banks:2006rx,Banks:2016taq,Susskind:2021dfc}, and the de Sitter static patch behaves as a holographic quantum mechanical system whose degrees of freedom are localized at the horizon. 

It is natural to wonder how well this viewpoint holds up in higher and lower dimensions and when quantum backreaction is accounted for. Here we compare and contrast the central points of \cite{Susskind:2021dfc} for classical de Sitter black holes with the qSdS solution. Importantly, we will see the entropy deficit of a quantum SdS black hole takes on a similar form as its classical four-dimensional counterpart, however, where the classical Bekenstein-Hawking entropy is replaced by the generalized entropy. Moreover, using the fact that the generalized entropy is a linear function of the mass, we compute the nucleation rate using the method of constrained instantons, extending \cite{Morvan:2022ybp} to the case when the backreaction of quantum fields is included.


\subsection{Entropy deficit}

Consider first the case of classical black holes. Let $S_{0}$ denote the entropy of four-dimensional de Sitter space in the static patch
\beq S_{0}=\frac{4\pi R_{4}^{2}}{4G_{4}}\;,\eeq
where $R_{4}$ is the length scale of $\text{dS}_{4}$. This entropy is understood to be the entropy when de Sitter space is in thermal equilibrium, such that $S_{0}$ is maximized at a given average energy. Fluctuations may arise and shrink the cosmological horizon so that the entropy becomes less than $S_{0}$. Denote this smaller entropy by $S_{1}$. The probability $\mathcal{P}$ of such fluctuations depends on the entropy deficit $\Delta S=S_{0}-S_{1}$, \beq\mathcal{P}\sim e^{-\Delta S}\,.\eeq
For example, consider a fluctuation in which a small black hole with horizon radius $r_{+}$ and mass $M$ appears in $\text{dS}_{4}$. The geometry is given by the standard four-dimensional Schwarzschild-de Sitter solution
\beq ds^{2}=-f(r)dt^{2}+f^{-1}(r)dr^{2}+r^{2}d\Omega_{2}^{2}\;,\quad f(r)=1-\frac{r^{2}}{R_{4}^{2}}-\frac{2MG_{4}}{r}\;,\label{eq:sds4line}\eeq
where `small' here implies $r_{+}\ll R_{4}$. To lowest order in $M$, the horizon radius is 
$r_{+}=R_{4}-MG_{4}$, and the subsequent entropy $S$ is 
\beq S
=S_{0}-2\pi R_{4}M=S_{1}\;,\eeq
where in the last equality we identified the entropy $S=S_{1}$ since including a black hole lowers the entropy of de Sitter space. The entropy deficit $\Delta S$ is then 
\beq \Delta S=2\pi R_{4}M=\sqrt{S_{0}s}\;,\label{eq:entdefsuss}\eeq
where $s=4\pi M^{2}G_{4}$ is the entropy of a four-dimensional flat space Schwarzschild black hole.

More generally, for $d$-dimensional de Sitter black holes, where now
\beq f(r)=1-\frac{r^{2}}{R_{d}^{2}}-\frac{16\pi G_{d}M}{(d-2)\Omega_{d-2}r^{d-3}}\;,\quad \Omega_{d-2}=\frac{2\pi^{(d-1)/2}}{\Gamma[(d-1)/2]}\;,\eeq
and $R_{d}$ is the $d$-dimensional de Sitter radius, the entropy deficit (\ref{eq:entdefsuss}) for $s\ll S_{0}$ becomes 
\beq \Delta S=2\pi R_{d}M=\left(\frac{d-2}{2}\right)S_{0}^{\frac{1}{d-2}}s^{\frac{d-3}{d-2}}\;.\label{eq:entdefddim}\eeq
Here $s$ is now the entropy of a $d$-dimensional (flat space) Schwarzschild black hole. 

Expressing the entropy deficit as $\Delta S=\sqrt{S_0s}$ is interesting in that it depends on the entropies of two systems: the black hole and cosmological horizons. 
Motivated by \cite{Banks:2006rx}, ref.~\cite{Susskind:2021dfc} argued the entropy deficit (\ref{eq:entdefsuss}) is reproduced by M(atrix) theory \cite{Banks:1996vh}, such that the holographic degrees of freedom of de Sitter space may be represented by $N\times N$ Hermitian matrices,
%
%
and the entropy deficit (\ref{eq:entdefsuss}) follows in a straightforward way (up to an overall factor of two).\footnote{The entropy deficit for $d$-dimensional systems (\ref{eq:entdefddim}) may also be given a matrix model interpretation, however, doing so requires the entropy per degree of freedom to depend on the size of $N\times N$ matrix, $\sigma(N)\sim N^{-(d-4)/(d-3)}$ \cite{Susskind:2021dfc}, which we see is potentially problematic for $d=3$.}


\subsection*{Matrix model description of quantum black holes}

We see that the entropy deficit for classical black holes \eqref{eq:entdefddim} does not allow us to consider the three-dimensional case. In the limit of $d=3$, one has $\Delta S=S_0/2$, where $S_{0}$ is the entropy of empty $\text{dS}_{3}$. It is clear that the dependence on $s$ vanishes since there are no classical black holes in dS$_3$. However, our construction allows us to have such black holes, when they are immersed into quantum fields. Two questions naturally arise: what is the entropy deficit when backreaction is accounted for, and does it have a similar matrix model interpretation?


Clearly, the entropy deficit of nucleating a quantum SdS black hole in (quantum) de Sitter will account for the entropy of quantum fields, as they make a non-negligible contribution to the overall state. Heuristically, we expect the entropy deficit to be related to a difference in generalized entropies, 
\beq \Delta S^{(3)}_{\text{gen}}=S^{(3)}_{\text{gen},0}-S^{(3)}_{\text{gen}}\;,\label{eq:entdefgen}\eeq
where $S^{(3)}_{\text{gen},0}$ is the entropy of the $\text{dS}_{3}$ cosmological horizon including quantum fields (the analog of $S_{0}$), and $S^{(3)}_{\text{gen}}$ denotes the entropy of nucleating a small quantum SdS black hole (the analog of $S_{1}$). This deficit may be computed explicitly (see App.~\ref{app:susskind} for details), however, using braneworld holography we may deduce what to expect. 

Holographically, the three-dimensional generalized entropy of the backreacted geometry on the brane is identified with the classical four-dimensional Bekenstein-Hawking entropy of the bulk solution, 
\beq S_{\text{BH}}^{(4)} \equiv S_{\text{gen}}^{(3)}. \eeq
Without performing an explicit computation, then, we expect the generalized entropy deficit (\ref{eq:entdefgen}) will exhibit the same behavior as the classical four-dimensional entropy deficit (\ref{eq:entdefsuss}),
\beq \Delta S^{(3)}_{\text{gen}} \sim \sqrt{S_{\text{gen},0} s_{\text{gen}}}, \label{eq:sgenSdefv1}\eeq 
where $s_{\text{gen}}$ the generalized entropy for a black hole in three-dimensional flat space. The `$\sim$' denotes the fact that we expect the entropy deficit may depend on a proportionality factor which depends on the relevant scales, namely $\ell$ and $R_{3}$; indeed, we find the relevant factor is $(1+R_{3}/\ell)^{1/2}$ (see App.~\ref{app:susskind}). Comparing the deficit (\ref{eq:sgenSdefv1}) to the four-dimensional entropy deficit (\ref{eq:entdefsuss}), suggests the holographic degrees of freedom of quantum de Sitter space may likewise be represented by $N\times N$ Hermitian matrices, modulo the overall factor. 

Similarly, using the fact that the relation between generalized entropy and mass for holographic conformal fields is the same as for a classical horizon in one more dimension,\footnote{We are ignoring the effects of possible additional compact dimensions, $AdS\times \mathcal{M}$.} it is natural to extend this modified deficit to arbitrary spacetime dimension,
\beq \Delta S_{\text{gen}}=\mathcal{F}(d,\nu)S_{\text{gen},0}^{\frac{1}{d-1}}s_{\text{gen}}^{\frac{d-2}{d-1}}\;,\eeq
where $\mathcal{F}(d,\nu)$ is some constant factor which depends on the number of dimensions and the backreaction parameter in such a way so that one recovers the classical deficit (\ref{eq:entdefddim}). The lesson one can extract from this simple exercise is that quantum fields play a crucial role in computing the entropy deficit. Consequently, backreaction due to quantum fields will affect nucleation rates, as we now describe.

\subsection{Nucleation rate}


Entropy deficits play a key role in characterizing black hole nucleation rates. For example, consider the entropy deficit of a classical dS black hole of mass $M$. 
\beq \Delta S=S_{0}-S_{\text{tot}}=\frac{S_{0}}{3}(1-y^{2})\;.\label{eq:entdefv2}\eeq
Here $S_{\text{tot}}$ is sum of the entropies $S_{h,c}=\frac{4\pi r_{h,c}^{2}}{4G_{4}}$, and $y\equiv(r_{c}-r_{h})/R_{4}$ is a dimensionless length distinguishing the black hole and cosmological horizons in SdS. For $0<M<M_{\text{N}}$, $y\in[-1,1]$, where $y=0$ corresponds to the Nariai limit.\footnote{Each range $y\in[-1,0]$ and $y\in[0,1]$ can be viewed as physically distinct configurations: for $y<0$, the black hole horizon grows while the cosmological horizon shrinks, until eventually $r_{h}=r_{c}$ (the Nariai limit), while for $y>0$ the horizons swap roles, such that $r_{c}>r_{h}$. 
}
In  \cite{Susskind:2021dfc} it is argued that the probability to nucleate a  black hole of \emph{fixed} mass follows from integrating $\Delta S$ with respect to $y$
\beq \mathcal{P}\sim \int_{0}^{1}\frac{d^{4}y}{R_{4}^{2}}e^{-\Delta S(y)}=
\left(\frac{3}{S_{0}}-\frac{9}{S_{0}^{2}}\right)+\frac{9}{S_{0}^{2}}e^{-S_{0}/3}\;.
\label{eq:nucratesuss}\eeq
Expressing $S_{0}=\pi R_{4}^{2}/G_{4}$, one sees the term between brackets is perturbative in $G_{4}$ and accounts for the non-universal microphysics of small black holes. Meanwhile, the second term is non-perturbative in $G_{4}$ but is universal, representing a saddle point when the integrand is at $y=0$, and thus characterizes a contribution from the Nariai geometry.


While it is intuitive to motivate the nucleation rate (\ref{eq:nucratesuss}) in terms of the entropy deficit, the above calculation is unsatisfactory for two reasons. First, the parameter $y$ is not well physically motivated and expressing $\Delta S$ in terms of $y$ for higher dimensional black holes is technically challenging. Second, 
the nucleation rate above assumes a fixed mass, but one is generally interested in the nucleation of arbitrary mass black holes. Further, black hole nucleation may be naturally understood as a quantum tunneling process, analogous to bubble nucleation in vacuum decay \emph{viz} Coleman and de Luccia \cite{Coleman:1980aw}.

Therefore, below we advocate computing the nucleation rate via the difference between the on-shell Euclidean action of two different spacetimes. 
The benefit of this approach not only resolves the aforementioned points (reviewed briefly below), but also easily applies to the quantum black hole case, allowing us to interpret Euclidean qSdS as a ``constrained instanton", and show the nucleation rate is controlled by $\Delta S_{\text{gen}}$.


\subsection*{Nucleation of classical dS black holes}

In general, one encounters an immediate obstruction in applying the on-shell action method for de Sitter black hole nucleation. The reason is that the Euclideanized SdS background has two conical singularities, one for each horizon. Thus, while one conical singularity may be removed via an appropriate identification of the Euclideanized time coordinate $\tau=it$, 
there will remain a conical singularity.\footnote{Two special cases where there is only a single conical singularity include: (i) pure de Sitter space, where $M=0$, and where $\tau\sim \tau+\beta_{\text{GH}}$, and (ii) the Nariai geometry, where $\beta_{h}=\beta_{c}=\beta_{\text{N}}$, such that $\tau\sim \tau+\beta_{\text{N}}$.} However, for $d=4$ \cite{Gregory:2013hja} and $d\geq4$ \cite{Morvan:2022ybp}, it was shown the on-shell Euclidean action of the SdS solution is
\beq I_{\text{E},\text{SdS}}
=-(S_{h}+S_{c})=-S_{\text{tot}}\;.\label{eq:totalentIE}\eeq
This 
holds for an arbitrary periodicity $\beta$ of the Euclidean time. Now, an important feature of the total gravitational entropy (\ref{eq:totalentIE}) of the SdS solution is that, in any dimension, it is approximately a linear function of the mass $M$ with a negative slope for $M\in[0,M_{\text{N}}]$, 
\beq S_{\text{tot}}\approx S_{0}-(S_{0}-S_{\text{N}})\frac{M}{M_{\text{N}}}\;,\label{eq:linbehaviorS}\eeq
with $S_{0}$ being the usual entropy of the cosmological horizon in empty de Sitter space.

Heuristically then, the nucleation rate for a black hole of mass $M$ to spontaneously appear in empty de Sitter is given by 
\beq \mathcal{P}\sim e^{-(I_{\text{E},\text{SdS}}-I_{\text{E},\text{dS}})}\sim e^{-\Delta S}\;,\label{eq:probactionSdS}\eeq
where the second equality follows from using that $I_{\text{E},\text{SdS}}(M=0)=I_{\text{dS}}=-S_{0}$. 
The linear approximation  (\ref{eq:linbehaviorS}) resolves the technical challenge of using the parameter $y$ for higher dimensional black holes.

Importantly, the probability (\ref{eq:probactionSdS}) follows from a Euclidean path integral, where Euclidean SdS black holes represent constrained instantons \cite{Cotler:2020lxj,Morvan:2022ybp,Draper:2022xzl}. More carefully, it is natural to expect the nucleation rate to be described by instanton effects, particularly given the non-perturbative behavior expressed in (\ref{eq:nucratesuss}). However, it is not a standard instanton, i.e., it is not a solution to the classical Euclidean equations of motion. Rather, the Euclidean SdS geometry is 
a ``constrained instanton": a stationary point of the Euclidean action when a particular constraint is imposed, namely, fixing the mass.
Consequently, the probability rate of creating an arbitrary mass black hole in de Sitter is computed semi-classically via \cite{Morvan:2022ybp}
\beq \mathcal{P}\sim  \int_{0}^{M_{\text{N}}}d M e^{-\Delta S}\;.\label{eq:Pconstrained}\eeq
  Implementing the linear approximation (\ref{eq:linbehaviorS}), such that $\Delta S=(S_{0}-S_{\text{N}}) \frac{M}{   M_{\text{N}}}$, the integral (\ref{eq:Pconstrained}) may be precisely evaluated leading to 
\beq \mathcal{P}\sim \frac{M_{\text{N}}}{S_{0}-S_{\text{N}}}\left[1-e^{-(S_{0}-S_{\text{N}})}\right]  = \frac{M_{\text{N}}}{S_{0} /3}\left(1-e^{- S_{0}/3}\right)\;,\label{eq:Pconstrainedv2}\eeq
This is fairly different from
(\ref{eq:nucratesuss}), because we integrated $e^{-\Delta S}$   over $M$ instead of over  $y$. The pair creation rate \eqref{eq:Pconstrainedv2} has a constant contribution and a non-perturbative term coming from the Nariai instanton, but we note the non-perturbative contribution has a factor $3/S_0$ in front instead of $9/S_0^2$ as in (\ref{eq:nucratesuss}). The overall factor $M_\text{N}$ appears so that $\mathcal P$ has the dimensions of a probability rate per Hubble volume.

\begin{figure}[t!]
\begin{center}
\includegraphics[width=7.5cm]{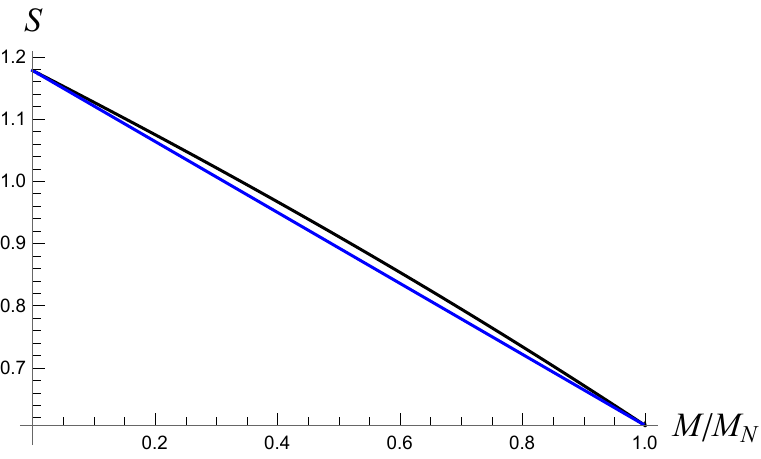}$\quad\;\;$\includegraphics[width=7.5cm]{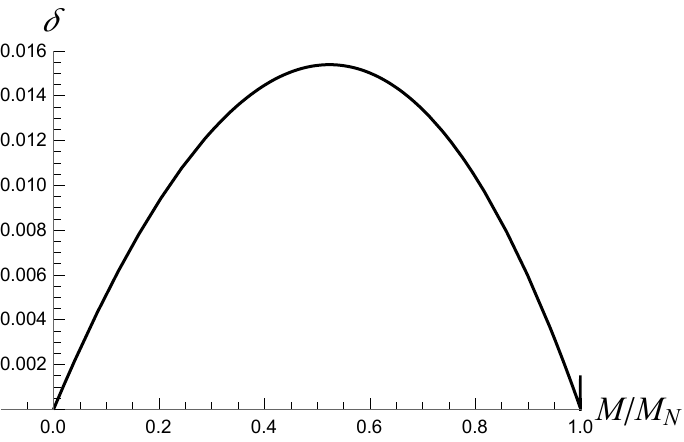}
\end{center}
\caption{\small \textbf{Left}: Total generalized entropy (black) and its linear fit (blue). \textbf{Right}: Difference $\delta$ between the total generalized entropy and the linear fit. Here we have set $\nu=1/3$.}
\label{fig:linfitent} 
\end{figure}

\subsection*{Nucleation of quantum dS black holes}

Given our holographic set-up, we expect the qSdS solution to likewise behave as a constrained instanton, such that the probability of nucleation is
\beq 
\begin{split}
    \mathcal{P}\sim& \int_{0}^{M_{\text{N}}}d M e^{-(I_{\text{E},\text{qSdS}}-I_{\text{E},\text{qdS}})}\approx   \int_{0}^{M_{\text{N}}}d Me^{-\Delta S_{\text{gen}}^{(3)}}\\
    &\approx\frac{M_{\text{N}}}{S^{(3)}_{\text{gen},0}-S^{(3)}_{\text{gen},\text{N}}}\left[1-e^{-(S^{(3)}_{\text{gen},0}-S^{(3)}_{\text{gen},\text{N}})}\right]\;,
\end{split}
\label{eq:PconstrainedqSdS}\eeq
analogous to (\ref{eq:Pconstrainedv2}).

The second equality technically follows from computing the on-shell Euclidean action of the effective three-dimensional theory (\ref{eq:totalact}).\footnote{Despite not knowing $I_{\text{CFT}}$ explicitly, this can be achieved since the effective gravitational contribution is known exactly (though perturbatively), and $I_{\text{CFT}}$ follows via subtracting the gravitational effective action from the known bulk action.} Meanwhile, the final equality follows because, as in the classical case, the total entropy generalized entropy 
is nearly linear in $M/M_{\text{N}}$ (Fig.~\ref{fig:Sgenwithtot}). Thus, we find the total generalized entropy is well approximated by the linear fit
\beq S^{(3),\text{tot}}_{\text{gen}}\approx S^{(3)}_{\text{gen},0}-(S_{\text{gen},0}^{(3)}-S^{(3)}_{\text{gen},\text{N}})\frac{M}{M_{\text{N}}}\;,\eeq
analogous to (\ref{eq:linbehaviorS}) (see Fig.~\ref{fig:linfitent}).

Therefore, the entropy deficit is approximately 
\beq \Delta S_{\text{gen}}^{(3)}\approx (S_{\text{gen},0}^{(3)}-S^{(3)}_{\text{gen},\text{N}}) \frac{M}{M_{\text N}}\;,\eeq
which leads to the final line in (\ref{eq:PconstrainedqSdS}).




\section{Comments on the holographic dual of de Sitter} \label{sec:commentsdSholo}

Whether de Sitter spacetimes have a holographic description remains one of the most important outstanding questions in quantum gravity. Inspired by AdS/CFT, multiple and distinct pictures of de Sitter holography have been proposed, including in particular \cite{Strominger:2001pn,Witten:2001kn,Maldacena:2002vr,Alishahiha:2004md,Banks:2005bm,Alishahiha:2005dj,Anninos:2011af,Dong:2018cuv,Gorbenko:2018oov,Leuven:2018ejp,Coleman:2021nor,Susskind:2021dfc, VanRaamsdonk:2021qgv,Araujo-Regado:2022gvw}. A distinguishing feature of each proposal is where the dual non-gravitational microscopic theory lives. On the one hand, the asymptotic fall-off of dS suggests a dual CFT should reside at asymptotic infinity, the natural analog of standard AdS/CFT, where the dual theory lives at the timelike conformal boundary. Alternatively, to better understand the thermodynamics of cosmological horizons, a natural place to put the dual microscopic theory may be on the (stretched) cosmological horizon, or a York-like ``boundary'' near the poles. While none of the proposals are fully satisfactory, the contrasting features of each make clear that de Sitter holography is an important open problem.

Given the recent successes of models exhibiting double holography \cite{Almheiri:2019hni,Chen:2020uac,Chen:2020hmv,Hernandez:2020nem}, it is natural to wonder whether this viewpoint has the ability to address the problem of de Sitter holography. 
That is, we can ask if the doubly-holographic perspective can shed some light on the nature of a tentative de Sitter dual. Until we find a top-down construction, it is an effective description and will not reveal the ultraviolet holographic degrees of freedom, but perhaps it can help with the question of where these degrees of freedom are located.

Let us first recall how the standard double holography works for AdS branes.  
%
%
\begin{figure}[t]
    \centering
    \includegraphics[width=.3\textwidth]{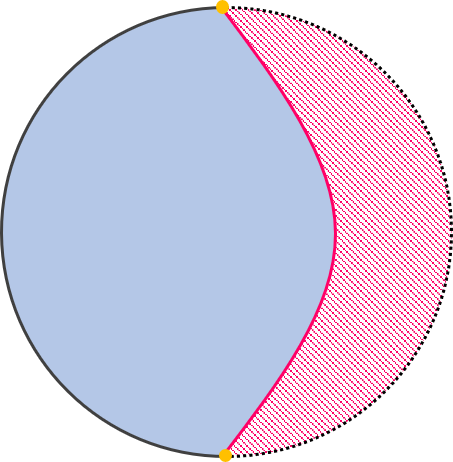}$\qquad\qquad\quad$ \includegraphics[width=.3\textwidth]{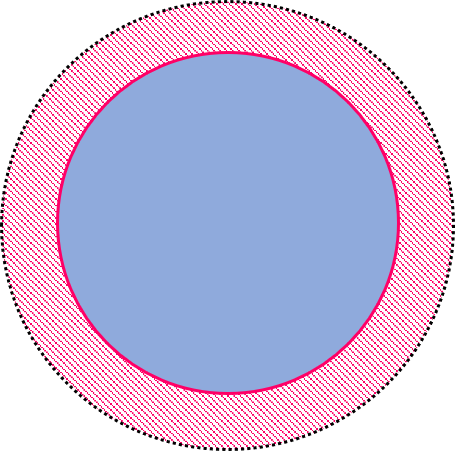}
    \caption{\small \textbf{Left}: An AdS brane embedded into a higher dimensional bulk. The shaded magenta region, including the dashed line of the AdS boundary, has been integrated out.  The two points in yellow (a circle, when rotated along the symmetry axis) represent two-dimensional defect CFTs, which are coupled to a bath CFT$_3$. The defect CFT is dual to the brane. \textbf{Right}: A timeslice of the dS brane embedded into a higher dimensional bulk. The blacked dashed line corresponds to the part of the boundary which has been integrated out, along with the shaded magenta region.}
    \label{fig:adsanddS}
\end{figure}
In this case, an AdS$_{d}$ brane hits the conformal boundary of a bulk AdS$_{d+1}$ spacetime at a $S^{d-2}$ that extends in time, as shown in Fig.~\ref{fig:adsanddS}. This sphere corresponds to a defect CFT$_{d-1}$, which interacts   with the bath CFT$_d$ living at the boundary. In the case of an AdS$_3$ brane, the defect CFT$_2$ lives on a circle. The reason why this picture is known as double holography is due to the nature of the brane. Namely, the brane dynamics consists of gravity with AdS$_d$ asymptotics, coupled to the same bath CFT$_d$ to which the defect CFT$_{d-1}$ is coupled. The CFT$_d$ on the brane together with its counterpart on the asymptotic boundary allow for a higher dimensional bulk. On the other hand, the AdS$_d$ asymptotics of the brane allow for dualization to a CFT$_{d-1}$, namely, the defect theory. Therefore, if one was to decouple the defect from the bath, one would simply obtain the usual AdS$_3$/CFT$_2$ setup. In other words, we can say that the defect CFT holographically describes the brane.

With a dash of speculation, we can claim that all branes can be sourced in terms of defect CFTs which couple to the bath CFT. What then does such a speculation entail for the de Sitter setup constructed here? 

Looking at a generic time slice of our setup, as shown in the right-hand side of Fig.~\ref{fig:adsanddS}, we see that the dS brane appears to be completely independent of the boundary CFT. In other words, it seems as if it does not reach the boundary. Of course, a single snapshot can be misleading, since the dS spacetime is described by an expanding hyperboloid, as shown in Fig.~\ref{fig:AdSwithdSbrane}. The hyperboloid hits the boundary again at two distinct spheres, but now these are moments at finite global time. We can easily see this if we transform the metric of AdS$_4$ in the dS$_3$ foliation, \eqref{eq:lineelementds3sec}, into  the conventional global form of AdS$_4$,
\begin{equation}
    ds^2 = -(1+R^2)dT^2+\frac{dR^2}{1+R^2}+R^2 d\Omega_2
\end{equation}
 (we set $\ell_4=1$ for simplicity), which is obtained from the coordinates in \eqref{eq:lineelementds3sec} by taking
\begin{equation}
    T=\arctan\left( \sqrt{1-\frac{r^2}{R_3^2}}\,\sinh\frac{t}{R_3}\,\tanh\sigma\right)\,,
\end{equation}
and
\begin{equation}
   R=\sinh\sigma\,\sqrt{ \left(1-\frac{r^2}{R_3^2}\right)\sinh^2\frac{t}{R_3} +1}\,.
\end{equation}
We see that, along a brane at finite $\sigma=\sigma_b$, the limits $t\to\pm\infty$ correspond to reaching the boundary $R\to\infty$ at finite global time $T\to \pm\pi/2$.

Taking the same perspective as in the case for AdS branes, we see that the analogue of defect CFTs is now played by two Euclidean CFTs, disconnected from the boundary point of view, but connected in the bulk through the brane. In other words, the Euclidean defect CFTs holographically describe the brane, or equivalently, the dual to the de Sitter brane is given by two Euclidean CFTs. Naturally, this picture hints at the dS/CFT construction of \cite{Strominger:2001pn, Witten:2001kn}. The main difference here lies in the fact that we have two states which source the brane, instead of the usual one-state preparation which fuels the dS phase. Regardless, the two views would seem to be, in terms of calculable observables, equivalent. 

A key advantage one might try to extract from our perspective is the possibility of utilizing double holography in our favor. Namely, one can use the higher-dimensional bulk as a means for computing relevant observables, instead of relying on somewhat ill-defined Euclidean CFT computations, as in the usual dS/CFT setup. However, one immediately comes to a halt.


The braneworld setup described here can be interpreted in terms of vacuum decay, as was done in \cite{Maldacena:2010un} (see also \cite{Barbon:2010gn,Barbon:2011ta}). Recall that vacuum decay from false vacuum to true vacuum \emph{\'{a} la} Coleman and de Luccia \cite{Coleman:1980aw} is understood as a tunneling process through the nucleation of a bubble of true vacuum which then expands and ``eats up'' the false vacuum. In our picture, the true vacuum is the AdS$_4$ bulk, the bubble is given by the dS brane, and the false vacuum can be seen as \textit{nothing}.\footnote{Alternatively, one can consider adding Minkowski patches to the sides of the de Sitter hyperboloid, in which case the false vacuum is simply given by Minkowski space.} Therefore, starting from some $t = 0$ slice in the middle of our hyperboloid, we are evolving towards a fully completed AdS vacuum.

\begin{figure}[t]
\begin{center}
\includegraphics[width=.5\textwidth]{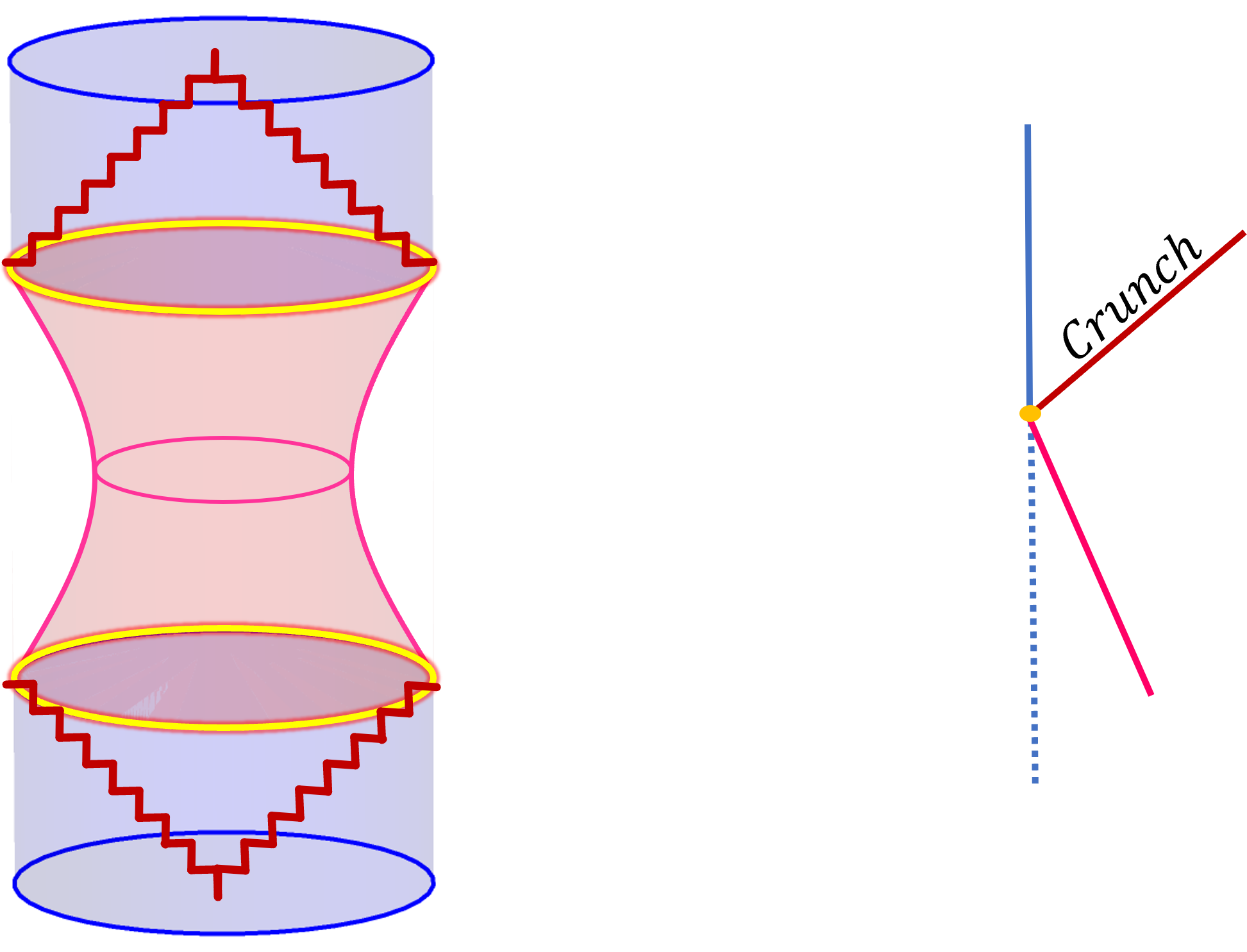}
\end{center}
\caption{\small A possible dS/CFT setup via braneworlds. \textbf{Left:} Bulk viewpoint. A bulk $\text{AdS}_{d+1}$ cylinder (in blue) has  a holographic $\text{CFT}_{d}$ dual living at the boundary. The $\text{dS}_{d}$ braneworld, with cutoff surfaces in magenta, lives inside $\text{AdS}_{d+1}$. One might envisage dualizing the $\text{dS}_{d}$ gravity with a Euclidean $\text{CFT}_{d-1}$ living at $\mathcal{I}^{-}$ and $\mathcal{I}^{+}$ of the de Sitter hyperboloid (yellow). Bulk quantum corrections are expected to drive the braneworld model to incur big bang and big crunch-like singularities (red), replacing the upper and lower Lorentzian cylinders. Alternatively, one could avoid the singularities altogether by preparing the `in' and `out' states on $\mathcal{I}^{-}$ and $\mathcal{I}^{+}$ via a Euclidean path integral with suitable sources turned on. This would entail replacing the cylinders by appropriate Euclidean submanifolds, closing the contour of the path integral \emph{\`{a} la} Schwinger-Keldysh \cite{Skenderis:2008dh,Skenderis:2008dg,Botta-Cantcheff:2015sav}. \textbf{Right:} The dual quantum mechanical picture, with time running upwards.}
\label{fig:dScftfig} 
\end{figure}

The vacuum decay as described in \cite{Coleman:1980aw} and \cite{Maldacena:2010un} leads to a Big Bang/Big Crunch singularity in the past and the future of the Euclidean CFT slices, respectively. One can understand the appearance of the singularity in the following sense: the dS brane is an accelerating brane, and as such it radiates at the quantum level. However, the brane time is infinite, since we are reaching the asymptotic infinities of the dS spacetime. Therefore, the amount of radiation such a brane will give off is infinite as well, creating a piling of rays at the future Cauchy horizon, as shown in Fig.~\ref{fig:dScftfig}. Thus, we see that the formation of the singularity is essentially the same phenomenon that enforces Strong Cosmic Censorship at the inner Cauchy horizon in charged or rotating black holes. It will also happen for the past Cauchy horizon.\footnote{One might wonder why such an event does not happen in usual AdS spacetimes, given that we can have the same dS slicing of empty AdS as well. The key difference lies in the amount of radiation one produces: in order to model the brane radiation, one would need to employ infinite shocks from the boundary in order to produce the same singular effect. Of course, such a state would be pathological to begin with, and so we have no (naturally formed) Big Bang and Big Crunch singularities.} Hence, it is not clear in what way we can exploit the doubly-holographic setup, since such a singularity might decouple the Euclidean CFTs from the rest of the Lorentzian CFT$_3$ bath.

Nevertheless, we can take the picture we obtained at face value. The duals claimed here arise as a result of a pure AdS/CFT construction. Therefore, we can see that the AdS/CFT construction itself hints that something like dS/CFT might supply an appropriate notion for a de Sitter dual (if one exists).\footnote{One can take this observation a step further (or back) and see a similar story holds for asymptotically flat branes. In that case, the dual would be given in terms of null defects coupled to the Lorentzian bath. There might be subtleties involving high energy shocks from the tip of the null defect, although it is not clear a singularity of the Coleman-de Luccia type would exist. We thank Jamie Sully for emphasizing this point.} One interesting thing to note is the necessity of correlating the boundary conditions associated to the Euclidean CFTs -- otherwise, we would not have an emergent geometry connecting the two theories. Usually, this would be interpreted as a problem for factorization. However, as in the case of low-dimensional AdS/CFT, one might resort to ensemble averaging of some sort in order to deem the correlated boundary conditions more natural. We plan to investigate this viewpoint in the future.

\section{Conclusion} \label{sec:disc}

While it is well known there are no classical black holes in $\text{dS}_{3}$, we have demonstrated quantum backreaction effects can generate a quantum black hole in $\text{dS}_{3}$. Our main analysis relied on $\text{AdS}_{4}/\text{CFT}_{3}$ braneworld holography, where the induced gravity action on a three-dimensional de Sitter brane admits a  quantum Schwarzschild de Sitter black hole, with backreaction due to the holographic $\text{CFT}_{3}$. A more conventional approach of computing the renormalized stress-tensor due to quantum matter, as carried out in Sec.~\ref{sec:Qstresstensec} (and App.~\ref{app:Qstressten}), points to a gravitationally attractive effect that is suggestive of black hole formation. However, this calculation is only reliable to linear order in the Planck length, while our holographic computation holds to all orders in the strength of backreaction (at planar order for the CFT), and therefore the presence of the horizon is well established.

We explored the thermodynamics of the qSdS solution, where we found the solution largely behaves like a classical, four-dimensional SdS black hole; however, importantly, all classical entropies are replaced by their generalized counterparts. In particular, we found it natural to interpret the quantum de Sitter black hole as a non-equilibrium constrained state of a thermodynamic system comprised of a bath (the quantum de Sitter background) and a subsystem (the quantum black hole). Equivalently, the qSdS black hole behaves as a constrained instanton where the probability of nucleating a black hole inside a quantum de Sitter background is controlled by the generalized entropy deficit. 

Our thermodynamic analysis also provides insights into the nature of the underlying microscopic degrees of freedom describing the quantum black hole and quantum de Sitter systems. 
The generalized entropy deficit  points to a matrix model description of quantum de Sitter space, where the holographic degrees of freedom of the cosmological horizon are represented by Hermitian matrices, in line with the classical SdS system in four dimensions~\cite{Susskind:2021dfc}. In fact, our inclusion of backreaction suggests a quantum generalization of the matrix model conjecture of \cite{Susskind:2021dfc}, even in higher dimensions. 

A particularly novel feature of our braneworld construction is that it leads to a framework for which we can study dS/CFT. Specifically, the de Sitter gravity theory on the brane is dual to a (defect) Euclidean CFT at $\mathcal{I}^{-}$ and $\mathcal{I}^{+}$. Advantageously, perhaps, via double-holography one can use the controlled setting of the bulk for computing observables of a typically ill-defined Euclidean CFT.

 Lastly, here we focused on quantum $\text{dS}_{3}$ black holes, a counterpart to the quantum $\text{AdS}_{3}$ black holes studied in \cite{Emparan:2020znc}. In the limit of large de Sitter radius $R_{3}$, but finite $\ell$, we recover quantum Schwarzschild black holes, with horizon radius $r_{h}=\mu \ell$, in asymptotically locally Minkowski space in three dimensions. These solutions were studied in \cite{Emparan:2002px,Emparan:2006ni}. 

There are a number of interesting research avenues worth pursuing, as we now describe.

\vspace{2mm}

\noindent \textbf{Rotating, charged quantum de Sitter black holes:} 
It is reasonably straightforward to include rotation, thus leading to rotating quantum de Sitter black holes in three dimensions. Similar to the rotating qBTZ solution \cite{Emparan:2020znc}, the starting point would be the rotating $\text{AdS}_{4}$ C-metric, however, with parameters tuned such that the brane has a positive cosmological constant. With the addition of rotation, it is expected that the   quantum Kerr-dS black hole will have a trifecta of correlated horizons (outer and inner black hole horizons and the cosmological horizon), leading to another extremal or ``lukewarm'' limit \cite{Romans:1991nq,Booth:1998gf,McInerney:2015xwa}, where the temperatures of the black hole and cosmological horizons coincide and are in principle distinct from the Nariai limit. One may also consider adding charge to the quantum black hole, starting from, for example, the charged $\text{AdS}_{4}$ C-metric. Although there is no need for counterterms for the Maxwell field in AdS$_4$, a Maxwell action is nevertheless generated on a brane at finite distance in the bulk \cite{inprep}, which modifies the geometry of the quantum-corrected black hole. On a de Sitter brane, lukewarm and cold charged instantons are expected.


\vspace{2mm}

 \noindent \textbf{Euclidean action, quasi-local thermodynamics, and stability of quantum de Sitter:} A first principles method for analyzing the thermodynamics of black holes is to directly compute the canonical partition function using a saddle-point approximation of the Euclidean gravitational path integral, \emph{\`a la} Gibbons and Hawking \cite{Gibbons:1976ue}. However, the standard treatment by Gibbons-Hawking suffers from ambiguities for de Sitter spacetime, since Euclidean de Sitter has no asymptotic boundary where a temperature may be specified to define the canonical ensemble (see \cite{Banihashemi:2022jys} for an in depth discussion on this point). Alternatively, one may adapt the quasi-local formalism of   York \cite{PhysRevD.33.2092} to de Sitter backgrounds and analyze quasi-local thermodynamics. This was recently accomplished in two-dimensional de Sitter JT gravity \cite{Svesko:2022txo}, where backreaction was accounted for exactly and the cosmological system was found to have a negative heat capacity and is thus thermodynamically unstable. It would be interesting to carry out a similar analysis for the quantum SdS solution, and see whether quantum de Sitter is thermodynamically stable. 
 

\vspace{2mm}

\noindent \textbf{Holographic complexity in quantum de Sitter space:} Models of double holography have recently been used to explore various proposals for information theoretic descriptions of quantum gravity. In particular, the `complexity=volume' and `complexity=action' conjectures have been analyzed in braneworld models \cite{Hernandez:2020nem}, where the braneworld gravity was given a holographic description in terms of a defect CFT. More recently, the effects of quantum backreaction were accounted for in these braneworld scenarios in the context of the quantum BTZ black hole \cite{Emparan:2021hyr}, with the `complexity=volume' being considerably more favorable over the `complexity=action' scenario. Thus far, however, little attention has been given to studying complexity in de Sitter space (see, \emph{e.g.}, \cite{Reynolds:2017lwq,Chapman:2021eyy,Jorstad:2022mls}). It would be very interesting to see whether the braneworld setup developed here could be used to efficiently study the aforementioned proposals in a quantum de Sitter background where backreaction effects are incorporated, analogous to the qBTZ analysis \cite{Emparan:2021hyr}. Doing so would require a better understanding of the relation between the dual $\text{CFT}_{3}$, and the defect $\text{ECFT}_{2}$ replacing the gravity on the $\text{dS}_{3}$ brane. Alternatively, our picture of de Sitter braneworlds, may, in principle, be naturally incorporated into the description of complexity in terms of holographic state preparation \cite{Pedraza:2021mkh,Pedraza:2021fgp,Pedraza:2022dqi}, where complexity may be understood as the minimum number of `Lorentzian threads' attached to unitaries preparing a tensor network state. From this perspective, the de Sitter hyperboloid would represent the time evolution of a dual boundary state prepared by a Euclidean path integral, replacing the standard Lorentzian AdS cylinder. Nonetheless, Lorentzian threads can just as easily extend through the hyperboloid, the number density of which is expected to capture a measure of complexity, leading to a possible connection to the tensor network construction of de Sitter space advocated in \cite{Bao:2017qmt}.

\vspace{2mm}

\noindent \textbf{Entanglement wedge islands and radiation entropy:} Braneworlds and models of double holography act as a useful arena to study the black hole information paradox, by computing the fine grained entropy of Hawking radiation using the `island rule' \cite{Chen:2020uac,Chen:2020hmv,Almheiri:2019psy}, an extremization prescription of semi-classical generalized entropy. Analogous information paradoxes arise in cosmological and de Sitter backgrounds \cite{Hartman:2020khs,Aalsma:2021bit,Kames-King:2021etp}, where quantum extremal surfaces and islands play a key role. Since the qSdS solution  uncovered here naturally incorporates the effect of backreaction, we have analytic control over the extremization of the generalized entropy on the brane to study detailed aspects of quantum extremal islands in dS. 

\vspace{2mm}

\noindent \noindent\section*{Acknowledgments}
We are grateful to Ahmed Almheiri, Jos\'e Barb\'on, Raphael Bousso, Adam Brown, Jaume Garriga, Ruth Gregory, Tom Hartman, Matthew Headrick, Christopher Herzog, Stefan Hollands, Kristan Jensen, Juan Maldacena, Alexey Milekhin, Yasunori Nomura,  Edgar Shaghoulian, Eva Silverstein, Jon Sorce, James Sully, Leonard Susskind, and Zhenbin Yang for discussions and useful correspondence. RE is supported by MICINN grant PID2019-105614GB-C22, AGAUR grant 2017-SGR 754, and State Research Agency of MICINN through the ‘Unit of Excellence Maria de Maeztu 2020-2023’ award to the Institute of Cosmos Sciences (CEX2019-000918-M). JFP is supported by the `Atracci\'on de Talento' program (2020-T1/TIC-20495, Comunidad de Madrid) and by the Spanish Research Agency (Agencia Estatal de Investigaci\'on) through the Grant IFT Centro de Excelencia Severo Ochoa No. CEX2020-001007-S, funded by MCIN/AEI/10.13039/501100011033. AS is supported by the Simons Foundation via \emph{It from Qubit: Simons Collaboration on quantum fields, gravity, and information}, and EPSRC. MT is supported by the European Research Council (ERC) under the European Union’s Horizon 2020 research and innovation programme (grant agreement No 852386). MV is supported by the Republic and canton of Geneva and the Swiss National Science Foundation, through Project Grants No. 200020-182513 and No. 51NF40-141869 The Mathematics of Physics (SwissMAP). AS and MV acknowledge the University of Barcelona for hospitality where this work was initiated. RE, JP, AS and MT acknowledge the Galileo Galilei Institute in Arcetri, Italy for its vibrant environment that allowed for fruitful  discussions. MT and MV thank the    participants of the Peyresq Physics workshop 2022 for useful interactions on this work.

\appendix

\section{Gravitational attraction from negative energy}  \label{app:negenergy}

It seems paradoxical that the negative Casimir energy created by a conical defect generates an attractive gravitational potential. However, this is an instance of a wider phenomenon that is present, only in reverse, in a much better known setup, so we will begin with it.

\subsection*{Apparent gravitational repulsion from positive energy}

Let us write the Reissner-Nordstr\"om solution as
\begin{equation}
    ds^2=-\left( 1-\frac{2{\mathcal M}(r)}{r}\right) dt^2+\frac{dr^2}{1-\frac{2{\mathcal M}(r)}{r}}+r^2 d\Omega_2\,,
\end{equation}
where
\begin{equation}\label{effMr}
    {\mathcal M}(r)=M-\frac{Q^2}{2r}\,.
\end{equation}
The function ${\mathcal M}(r)$ can be regarded as the ``effective mass'' that acts on a neutral test particle at radius $r$. We see that the electric field of the black hole decreases the gravitational force on such a particle, relative to a neutral black hole with the same mass $M$. Indeed, the acceleration of a neutral particle at fixed position (following an orbit of $\partial_t$) is smaller if we increase the charge $Q$ of the background geometry while keeping $M$ fixed. Thus, the electric field would seem to have a repulsive gravitational effect, despite the fact that its energy density,
\begin{equation}
    \rho_{el}=-T^t{}_t=\frac{Q^2}{2 r^4}\,,
\end{equation}
is positive. 

The correct interpretation, however, is different. The asymptotic mass $M$ measures the gravitational effect of \emph{all} the energy sources that are inside a sphere at infinity,\footnote{Readers who feel uneasy about the `energy  sources' of the non-linear Schwarzschild or Reissner-Nordstr\"om solutions can sidestep the issue by linearizing gravity and considering localized mass sources. Our arguments equally go through.} including also the electromagnetic field energy. Thus, if we put a test particle on the surface of a sphere at finite radius, then the electromagnetic energy that lies outside this sphere will not have any gravitational pull on the particle. When computing the actual mass that attracts the particle at $r$, the electromagnetic energy outside the sphere of radius $r$ must be subtracted from $M$, resulting in \eqref{effMr}.

Thus we see that the apparent repulsive gravitation of the electromagnetic field energy merely reflects a reduced attraction due to the lower energy that is enclosed as we move to spheres of smaller radii.

\subsection*{Gravitational attraction from negative Casimir energy in $2+1$ dimensions}

The previous argument easily explains how the negative Casimir energy density \eqref{casstress} 
\begin{equation}
   \rho_{Cas}=-\langle T^t{}_t \rangle =-\frac{\hbar F(M)}{8\pi r^3}
\end{equation}
gives rise to gravitational attraction. The only subtlety is that, in $2+1$ dimensions, a classical localized mass does not generate 
any gravitational attraction itself, only a deficit angle. This deficit, measured on circles at large radii, is related to the mass as in \eqref{eq:deficitangleintro}. In our quantum-corrected solutions, considering for simplicity the asymptotically locally flat limit $R_3\to\infty$, we could define an effective mass
\begin{equation}
    {\mathcal M}(r)=M +\frac{\hbar F(M)}{4 r}\,,
\end{equation}
which indicates that the quantum corrections enhance the gravitational effects at finite $r$, not only increasing the deficit angle at finite $r$, but also accelerating neutral particles towards smaller radii. 

We now know how to understand this attraction. The asymptotic mass $M$ measures the gravitational effect of all the energy enclosed in a circle at infinity, including the Casimir energy. The attraction at finite $r$ is a consequence of having less negative energy enclosed in a circle of radius $r<\infty$ than in one at infinity. What makes the effect perhaps more surprising is that in $2+1$ dimensions the leading asymptotic mass term $M$ does not result in any attraction, while the finite-$r$ correction does. But its origin is the same as we have seen above.

It can easily be seen that the explanation for the effect given in \cite{Soleng:1993yh} reduces to the argument that we have presented here, only in a more elaborate form (accounting for an explicit mass source near $r=0$), but which is possibly less transparent.

\section{Renormalized stress tensor of conical defect in $\text{dS}_{3}$}  \label{app:Qstressten}

Here we provide the details for computing the renormalized quantum stress-energy tensor of a conformally coupled massless scalar field to the Einstein-Hilbert action for a conical defect in $\text{dS}_{3}$. 
Our treatment below follows standard techniques using point-splitting, as described in, for instance, \cite{Steif:1993zv,Souradeep:1992ia,Casals:2016ioo}. 

\section*{Conical defect in $\text{dS}_{3}$}


 Static, circularly symmetric solutions to the Einstein-Hilbert action in three dimensions with pointlike matter sources  and a positive cosmological constant $\Lambda=+1/R_{3}^{2}$ may be parameterized as 
\beq ds^{2}=-\left(1-8G_{3}M-\frac{r^{2}}{R_{3}^{2}}\right)dt^{2}+\left(1-8G_{3}M-\frac{r^{2}}{R_{3}^{2}}\right)^{-1}dr^{2}+r^{2}d\phi^{2}\;,\label{eq:dS3conicalapp}\eeq
with $-\infty<t<\infty$, $0\leq r\leq\infty$, and the  angular coordinate $\phi$ is $2\pi$-periodic. The value of $M$ lends two distinct scenarios: (i) $M=0$ corresponds to pure $\text{dS}_{3}$ in static patch coordinates with a cosmological horizon $r_{c}=R_{3}$, and (ii) for $8G_{3}M<1$ we have the Schwarzschild-de Sitter (SdS) solution. Case (ii) does not describe a black hole in dS$_3$, but rather a conical defect with a cosmological horizon at $r_{c}=R_{3}\sqrt{1-8G_{3}M}$.  To see appreciate this point, define the parameter $\gamma^{2}\equiv1-8G_{3}M$, and perform the following coordinate rescaling,
\beq \tilde t=\gamma t\;,\quad \tilde r=\gamma^{-1}r\;,\quad \tilde \phi=\gamma\phi\;.\label{eq:rescalecoord}\eeq
The $\text{dS}_{3}$ line element (\ref{eq:dS3conicalapp}) becomes
\beq ds^{2}=-\left(1-\frac{\tilde r^{2}}{R_{3}^{2}}\right)d\tilde t^{2}+\left(1-\frac{\tilde r^{2}}{R_{3}^{2}}\right)^{-1}d\tilde r^{2}+\tilde r^{2}d\tilde \phi^{2}\;.\label{eq:conicaldS3}\eeq
This looks like $\text{dS}_{3}$ in static patch coordinates, however, with the notable difference that now the angular variable has a different periodicity: $\tilde \phi \sim\tilde \phi +2\pi\gamma$. Thus, this spacetime exhibits a conical defect, if $0 \le \gamma < 1$, with a deficit angle $\delta  = 2\pi(1-\gamma)$.
One may interpret this solution as a massive point particle sourcing the curvature, producing a curvature singularity as a   delta function source at the pole of the static patch of dS$_3$. In fact, a conical deficit at the north pole also induces  a conical deficit at the  south pole, since   timeslices of dS$_3$ are closed. Hence, conical dS$_3$ solutions have   two point particles, one at each pole. 

Finally, we want to compare the metric in \eqref{eq:dS3conicalapp} with the original metric found by Deser-Jackiw to describe two point particles in dS$_3$  \cite{Deser:1983nh}. By performing the following coordinate transformation
\begin{equation}
r= \frac{R_3 \gamma}{\cosh (\gamma  \ln z)} = \frac{2 \gamma R_3 }{z^\gamma + z^{-\gamma}}\,, \qquad \text{or} \qquad z^\gamma =\frac{   \gamma -\sqrt{\gamma^2 - r^2/R_3^2} }{r/R_3}\,,
\end{equation}
  the metric \eqref{eq:dS3conicalapp} of conical dS$_3$ turns into  
\begin{align}
    ds^2 &=   - \gamma^2 \tanh^2 (\gamma \ln z ) d t^2 + \frac{\gamma^2R_3^2  (dz^2 + z^2 d \phi^2)}{z^2 \cosh^2 (\gamma \ln z )} \,,\\
    &=- \gamma^2 \left ( \frac{z^\gamma - z^{-\gamma}}{z^\gamma + z^{-\gamma}} \right)^2 dt^2+  \frac{4 \gamma^2 R_3^2  (dz^2 + z^2 d \phi^2)}{z^2  \left ( z^\gamma + z^{-\gamma}\right)^2} \,.
\end{align}
This agrees with the metric which  Deser-Jackiw use for conical dS$_3$, see Eqs. (3.5) and (3.8) in \cite{Deser:1983nh}. For   $\gamma=1$ the spatial part   describes a round sphere in stereographic coordinates.


\section*{Adding a massless conformally coupled scalar field}


 Consider a massless scalar field $\Phi$ conformally coupled to the Einstein-Hilbert action in three dimensions (\ref{eq:confmassfieldEHact}), whose classical matter stress-energy tensor $T_{\mu\nu}$ is given by
\beq T_{\mu\nu}=\frac{3}{4}\nabla_{\mu}\Phi\nabla_{\nu}\Phi-\frac{1}{4}g_{\mu\nu}(\nabla\Phi)^{2}-\frac{1}{4}\Phi\nabla_{\mu}\nabla_{\nu}\Phi+\frac{1}{4}g_{\mu\nu}\Phi\Box\Phi+\xi G_{\mu\nu}\Phi^{2}\;,\label{eq:stresstensor}\eeq
where $G_{\mu\nu}$ is the Einstein tensor and $\xi=\frac{1}{8}$. When the background is maximally symmetric we have $G_{\mu\nu}=-g_{\mu\nu}\Lambda$.
Meanwhile the scalar field equation of motion is
\beq (\Box-\xi R)\Phi=0\;.\label{eq:scaleom}\eeq
Here $R=\frac{6}{R^{2}_{3}}$ when we are in $\text{dS}_{3}$.
Upon invoking (\ref{eq:scaleom}), it is straightforward to verify the classical stress-energy tensor (\ref{eq:stresstensor}) is traceless and conserved, $g^{\mu\nu}T_{\mu\nu}=\nabla^{\mu}T_{\mu\nu}=0$. 

The Green function $G_{\text{CdS}_{3}}(x,x')$ which solves the scalar field equation of motion (\ref{eq:scaleom}) for a conical defect in $\text{dS}_{3}$ may be computed using the method of images, analogous to conical $\text{AdS}_{3}$ \cite{Steif:1993zv}. Generically, the Green function $G(x,x')$ with transparent boundary conditions imposed  is \cite{Avis:1977yn,Lifschytz:1993eb}
\beq G(x,x')=\frac{1}{4\pi}\frac{1}{|x-x'|}\;,\eeq
where 
$|x-x'|\equiv\sqrt{(x-x')^{a}(x-x')_{a}}$ is the chordal or geodesic distance between $x$ and $x'$ in the four-dimensional embedding space $\mathbb{R}^{2,2}$. For pure $\text{dS}_{3}$, the embedding coordinates $x^{a}=(X_{1},X_{2},T_{1},T_{2})^{T}$ are
\beq T_{1}=\sqrt{r^{2}-R_{3}^{2}}\cosh(t/R_{3})\;,\quad T_{2}=\sqrt{r^2-R_{3}^{2}}\sinh(t/R_{3})\;,\quad X_{1}=r\cos\phi\;,\quad X_{2}=r\sin\phi\;.\label{eq:embeddingcoorddS3}\eeq
It is easy to verify
\beq -T_{1}^{2}+T_{2}^{2}+X_{1}^{2}+X_{2}^{2}=R_{3}^{2}\;,\eeq
and
\begin{equation}
\begin{aligned}
    ds^{2}&=-dT_{1}^{2}+dT_{2}^{2}+dX_{1}^{2}+dX_{2}^{2}\\&=-\left(1-\frac{r^{2}}{R_{3}^{2}}\right)dt^{2}+\left(1-\frac{r^{2}}{R_{3}^{2}}\right)^{-1}dr^{2}+r^{2}d\phi^{2}\;.
\end{aligned}  
\end{equation}
Then, 
\beq |x-x'|=\left[2R_{3}^{2}+2\sqrt{r^{2}-R_{3}^{2}}\sqrt{r'^{2}-R_{3}^{2}}\cosh\left(\frac{t-t'}{R_{3}}\right)-2rr'\cos(\phi-\phi')\right]^{1/2}\;.\eeq
Further, one can show   
\beq \left(\Box-\frac{3}{4R_{3}^{2}}\right)G_{\text{dS}_{3}}(x,x')=0\;,\eeq
when $x\neq x'$, and where $G_{\text{dS}_{3}}(x,x')$ refers to the Green function with respect pure $\text{dS}_{3}$ in static patch coordinates.


One may construct the Green function $G_{\text{CdS}_{3}}(x,x')$ for the conical defect spacetime (\ref{eq:conicaldS3}) via the method of images. That is, one uses the fact that the conical defect spacetime corresponds to discrete identifications of $\text{dS}_{3}$.  Specifically,  analogous to the $\text{AdS}_{3}$ case (see, \emph{e.g.}, \cite{Casals:2016ioo}), identified points are related by an element $H\in SO(1,3)$ on the embedding space coordinates  (\ref{eq:embeddingcoorddS3}), except where $\phi\sim\phi+2\pi\gamma$, with $\gamma\equiv 1/N$ for some positive integer $N$,
\beq H=\begin{pmatrix} \cos(2\pi\gamma)&\sin(2\pi\gamma)&0&0\\-\sin(2\pi\gamma)&\cos(2\pi \gamma)&0&0\\ 0&0&1&1\\0&0&0&1\end{pmatrix}\;.\label{eq:rotmatrix}\eeq
The Green function $G_{\text{CdS}_{3}}(x,x')$ for the conical defect spacetime then follows from the image sum
\beq G_{\text{CdS}_{3}}(x,x')=\sum_{n=-\infty}^{\infty}G_{\text{dS}_{3}}(x,H^{n}x')=\frac{1}{4\pi}\sum_{n\in\mathbb{Z}}\frac{1}{|x-H^{n}x'|}\;,\label{eq:imagesum}\eeq
with
\beq |x-H^{n}x'|=\left[2R_{3}^{2}+2\sqrt{r^{2}-R_{3}^{2}}\sqrt{r'^{2}-R_{3}^{2}}\cosh\left(\frac{t-t'}{R_{3}}\right)-2rr'\cos\left(\phi-\phi'+\frac{2\pi n}{N}\right)\right]^{1/2}\;.\eeq
Crucially, in the conical defect spacetime, the infinite sum becomes a finite sum, 
\beq G_{\text{CdS}_{3}}(x,x')=\frac{1}{4\pi}\sum_{n=0}^{N-1}\frac{1}{|x-H^{n}x'|}\;,\label{eq:GreenfuncdS3}\eeq
which follows from the fact there exist only a finite number $N$ of geodesics connecting two points on a cone \cite{Matschull:1998rv}. Upon a Wick rotation $L= iR_{3}$, one recovers the scalar field Green function in conical $\text{AdS}_{3}$ \cite{Casals:2016ioo}.

\section*{Quantum stress tensor for a conical defect in $\text{dS}_{3}$}

We can now obtain the renormalized quantum stress tensor $\langle T_{\mu\nu}\rangle$ from $G(x,x')$ using the point-splitting method \cite{Christensen:1976vb,Wald:1978pj,Steif:1993zv,Souradeep:1992ia,Casals:2016ioo}. Specifically, 
\beq \langle T_{\mu\nu}(x)\rangle=\lim_{x'\to x}\left(\frac{3}{4}\nabla^{x}_{\mu}\nabla^{x'}_{\nu}G-\frac{1}{4}g_{\mu\nu}g^{\alpha\beta}\nabla^{x}_{\alpha}\nabla^{x'}_{\beta}G-\frac{1}{4}\nabla^{x}_{\mu}\nabla^{x}_{\nu}G+\frac{1}{16R_{3}^{2}}g_{\mu\nu}G\right)\;,\label{eq:qustresstensorads3}\eeq
where $G(x,x')=G_{\text{CdS}_{3}}(x,x')$ is the Green function (\ref{eq:GreenfuncdS3}), the metric $g_{\mu\nu}$ is a function of spacetime point $x$, $\nabla_{\mu}^{x}$ denotes a covariant derivative with respect to point $x$, and $\nabla_{\mu}^{x'}$ denotes a derivative with respect to the point $x'$. Moreover, the limit $x\to x'$ is the coincident limit, which amounts to evaluating the resulting expression at $x'=x$.  Note that while normally the renormalization of the stress tensor is difficult, here one simply subtracts off the $n=0$ term in the image sum in the coincident limit; indeed, the $n=0$ term includes the divergent contribution.

To evaluate each component of the renormalized stress tensor in the conical defect background, we recognize $G(x,x')$ is a symmetric biscalar, while its covariant derivatives are examples of bitensors. Consequently, one invokes a generalization of Synge's theorem for bitensors developed by Christensen \cite{Christensen:1976vb} (also see Eq. (54) of \cite{Herman:1995hm}):
\beq \lim_{x'\to x}(\nabla^{x'}_{\mu}A_{\alpha_{1}})=\nabla^{x}_{\mu}\lim_{x'\to x}(A_{\alpha_{1}})-\lim_{x'\to x}(\nabla^{x}_{\mu}A_{\alpha_{1}})\;,\label{eq:syngerule}\eeq
where $A_{\alpha_{1}}$ is a bivector with equal weight at both $x$ and $x'$, whose coincidence limit exists. Consequently, applying Synge's rule (\ref{eq:syngerule}) to the quantum stress tensor (\ref{eq:qustresstensorads3}) we have:
\begin{align}
 \langle T_{\mu\nu}(x)\rangle&=\frac{3}{4}\left[\nabla_{\nu}^{x}\lim_{x'\to x}(\nabla^{x}_{\mu}G)-\lim_{x'\to x}(\nabla^{x}_{\nu}\nabla^{x}_{\mu}G)\right]-\frac{1}{4}g_{\mu\nu}g^{\alpha\beta}\left[\nabla^{x}_{\beta}\lim_{x'\to x}(\nabla_{\alpha}^{x}G)-\lim_{x'\to x}(\nabla^{x}_{\beta}\nabla_{\alpha}^{x}G)\right] \nonumber \\
&+\lim_{x'\to x}\left(-\frac{1}{4}\nabla^{x}_{\mu}\nabla^{x}_{\nu}G+\frac{1}{16R_{3}^{2}}g_{\mu\nu}G\right)\;.
\label{eq:qustresstensorsynge}
\end{align}
Evaluating this in the conical defect spacetime (\ref{eq:dS3conicalapp}), we find all off-diagonal components vanish, 
leaving only non-zero diagonal contributions (\ref{casstress}) with form factor $F(\gamma)$ (\ref{eq:Tmunuads3app})



\section{Bulk dual of a CFT in conical $\text{dS}_{3}$} \label{app:bulkdualCFT}


Here we show the bulk dual description of a holographic CFT in conical $\text{dS}_{3}$ is equal to a double Wick rotation of the hyperbolic $\text{AdS}_{4}$ black hole. To see this, consider the limit of vanishing backreaction $\ell\to 0$ of the $\text{AdS}_{4}$ C-metric (\ref{eq:AdS4Ccoord}), 
\beq ds^{2}=\frac{\ell^{2}}{x^{2}r^{2}}\left[-\left(1-\frac{r^{2}}{R_{3}^{2}}\right)dt^{2}+\left(1-\frac{r^{2}}{R_{3}^{2}}\right)^{-1}dr^{2}+r^{2}\left(G^{-1}(x)dx^{2}+G(x)d\phi^{2}\right)\right]\;,\label{eq:AdS4l0app}\eeq
where $G(x)=1-x^{2}-\mu x^{3}$. In the limit $\ell\to0$, the $\text{AdS}_{4}$ cosmological constant yields $\ell_{4}=\ell$. Clearly, along boundary $x=0$, where $G(x)=1$, the above geometry is conformally equivalent to conical $\text{dS}_{3}$. Under the following double Wick rotation,
\beq t = -iR_{3}\Phi\;,\quad r= \frac{R_{3}}{\cosh u}\;,\quad x=\frac{1}{\hat{\rho}}\;,\quad \phi=-i\hat{T}\;,\eeq
with $u\in\mathbb{R}_{+}$ and $\Phi\in[0,2\pi]$, the line element (\ref{eq:AdS4l0app}) becomes
\beq ds^{2}=\ell_{4}^{2}\left[-\left(\hat{\rho}^{2}-1-\frac{\mu}{\hat{\rho}}\right)d\hat{T}^{2}+\left(\hat{\rho}^{2}-1-\frac{\mu}{\hat{\rho}}\right)^{-1}d\hat{\rho}^{2}+\hat{\rho}^{2}(du^{2}+\sinh^{2}(u)d\Phi^{2})\right]\;.\eeq
Observe that when taking the limit $\ell\to 0$ with $R_3$ fixed, we are also sending $\ell_4\to 0$, so the entire metric shrinks to zero size. We can nevertheless effectively blow up the metric to finite size again by rescaling
\beq \hat{\rho}=\frac{\rho}{\ell_{4}}\;,\quad \hat{T}=\frac{T}{\ell_{4}}\;,\quad \mu=\frac{2m G_{4}}{\ell_{4}}\;.\eeq
Then we recover the line element for the hyperbolic (or topological) $\text{AdS}_{4}$ black hole \cite{Birmingham:1998nr},
\beq ds^{2}=-f(\rho)dT^{2}+f^{-1}(\rho)d\rho^{2}+\rho^{2}(du^{2}+\sinh^{2}(u)d\Phi^{2})\;,\quad f(\rho)=\frac{\rho^{2}}{\ell_{4}^{2}}-1-\frac{2mG_{4}}{\rho}\;.\eeq
Here the parameter $m$ is related to the ADM mass $M$ via
\beq M=\frac{m\omega_{2}}{4\pi}=\frac{\omega_{2}}{8\pi G_{4}}\rho_{+}\left[\left(\frac{\rho_{+}}{\ell_{4}}\right)^{2}-1\right]\;,\eeq
where $\rho=\rho_{+}$ is the location of the horizon, and $\omega_{2}=2\pi\int_{0}^{\infty}\sinh(u)du$ is the volume of hyperbolic space with unit radius. The Bekenstein-Hawking entropy and temperature, meanwhile, are
\beq S_{\text{BH}}^{(4)}=\frac{\omega_{2}\rho_{+}^{2}}{4G_{4}}\;,\quad T_{\text{H}}=\frac{1}{4\pi\rho_{+}}\left(\frac{3\rho_{+}^{2}}{\ell_{4}^{2}}-1\right)\;.\eeq
This gravitational entropy is generally divergent due to the infinite extent of the $\mathbb{H}^{2}$ hyperbolic space $\omega_{2}$, in accordance with the entanglement entropy of the dual CFT, and may be regulated by introducing a cutoff. 

\section{Entropy deficit of small quantum dS black hole}
\label{app:susskind}



Consider the generalized entropy (\ref{eq:Sgen}) of the cosmological horizon evaluated about $r_{+}=R_{3}-\frac{\mu\ell}{2}$ at small $\mu$, where we keep $\ell$ fixed, but $\nu$ is still assumed to be small. 
Then, 
\beq S_{\text{gen}}^{(3)}\approx \frac{\pi R_{3}^{2}}{2\mathcal{G}_{3}(\ell+R_{3})}-\frac{\pi R_{3}\mu\ell}{2\mathcal{G}_{3}(\ell+R_{3})}+\mathcal{O}(\mu^{2})\;,\label{eq:Sgen3smallmu}\eeq
where we used $x_{1}\approx 1$ at small $\mu$.
In the limit $\ell\to0$, we find $S^{(3)}_{\text{gen}}=\frac{\pi R_{3}}{2 G_{3}}$, the entropy of empty $\text{dS}_{3}$. The entropy (\ref{eq:Sgen3smallmu}) is the analog of the classical entropy $S_{1}$ and is less than the maximum entropy $S^{(3)}_{\text{gen},0}=\frac{\pi R_{3}^{2}}{2\mathcal{G}_{3}(\ell+R_{3})}$, the entropy of the cosmological horizon of $\text{dS}_{3}$ including matter field fields (\ref{eq:SgenqdS}). The entropy deficit $\Delta S_{\text{gen}}^{(3)}$ is therefore,
\beq \Delta S_{\text{gen}}^{(3)}=S_{\text{gen},0}^{(3)}-S_{\text{gen}}^{(3)}\approx\frac{\pi R_{3}\mu\ell}{2\mathcal{G}_{3}(\ell+R_{3})} \approx 2\pi R_{3}M\;,\label{eq:entdeficit3d}\eeq
where in the last equality we used that $M\approx \frac{\mu\ell}{4\mathcal{G}_{3}(\ell+R_{3})}$ in the limit $r_{+}=R_{3}-\frac{\mu\ell}{2}$ for small $\mu$. 

We observe the deficit of the generalized entropy for the qSdS solution is precisely of the same form as the classical entropy deficit (\ref{eq:entdefsuss}), however, here the deficit vanishes in the case of vanishing backreaction. Note that had we instead considered a conical deficit in three-dimensional de Sitter space, the form of the entropy deficit would be the same, where the mass $M$ would be identified with the `mass' of the conical defect.

Further, we may write the deficit as
\beq \Delta S_{\text{gen}}^{(3)}=S_{\text{dS}_{3}}(2\mathcal{G}_{3}M)\;,\eeq
with $S_{\text{dS}_{3}}=\frac{2\pi R_{3}}{2\mathcal{G}_{3}}$. This is the analog of the classical four-dimensional result (\ref{eq:entdefsuss}). Alternatively, the generalized entropy of a small black hole $s_{\text{gen}}$ is proportional to $M^{2}$. To see this, note that the black hole horizon of our qSdS black hole increases linearly in $\mu\ell$. This follows from expanding (\ref{eq:R3muellrcrh}) about $r_{h}=0$, such that at leading order,
\beq R_{3}^{2}\approx r_{c}^{2}\left(1+\frac{r_{h}}{r_{c}}\right)\;,\quad \mu\ell\approx r_{h}\;.\eeq
Then, let $s_{\text{gen}}$ denote the value of $S_{\text{gen}}^{(3)}$ evaluated at $r_{+}\approx \mu\ell$, expanded about small $\mu$, 
\beq s_{\text{gen}}\equiv S_{\text{gen}}^{(3)}|_{r_{+}=\mu\ell}\approx \frac{\ell\pi\mu^{2}}{2\mathcal{G}_{3}}\;.\eeq
Notice $s_{\text{gen}}$ vanishes in the limit of zero backreaction, as expected. Further,  the mass $M$ \eqref{eq:mass3} goes as
\beq M|_{r_{+}=\mu\ell}\approx \frac{\mu}{4\mathcal{G}_{3}}\;,\eeq
such that 
\beq \frac{s_{\text{gen}}}{8\pi \mathcal{G}_{3}\ell}=M^{2}\;.\eeq
Consequently, we may alternatively recast the entropy deficit (\ref{eq:entdeficit3d}) as
\beq
\begin{split}
\Delta S^{(3)}_{\text{gen}}&=
\sqrt{(1 + R_3 / \ell )S_{\text{gen},0}^{(3)}s_{\text{gen}}} \;.
\end{split}
\label{eq:sgenSdef}\eeq
which has a similar form as the four-dimensional deficit (\ref{eq:entdefsuss}).

\bibliography{qdSrefs}

\end{document}